\documentclass[pra,aps,superscriptaddress,amssymb,amsmath,reprint,noeprint,floatfix]{revtex4-2}
\usepackage{natbib}
\usepackage{braket}
\usepackage{bm}
\usepackage{graphicx}
\usepackage{hyperref}
\usepackage{xcolor}
\hypersetup{bookmarksnumbered}

\newcommand{\EqnOne}{
    \begin{equation}
    \begin{aligned}
        &\mathbf{A}=\frac{1}{|\sqrt{\mathbf{E} \cdot \mathbf{E}}|} \operatorname{Re}\left\{\mathbf{E} \sqrt{\mathbf{E}^{*} \cdot \mathbf{E}^{*}}\right\},\\
        &\mathbf{B}=\frac{1}{|\sqrt{\mathbf{E} \cdot \mathbf{E}}|} \operatorname{Im}\left\{\mathbf{E} \sqrt{\mathbf{E}^{*} \cdot \mathbf{E}^{*}}\right\}.
    \end{aligned}
        \label{eqn:1}
    \end{equation}
}

\newcommand{\EqnTwo}{
    \begin{equation}
    I_{\mathrm{pl}}=\frac{1}{4 \pi} \oint d \phi, \quad I_{\mathrm{ph}}=\frac{1}{2 \pi} \oint \nabla \varphi \cdot d \mathbf{r}.
        \label{eqn:2}
    \end{equation}
}

\newcommand{\EqnThree}{
    \begin{equation}
    \mathbf{E}_{\mathrm{s}}=\sum_{n, m} i E_{m n}\left(a_{m n} \mathbf{N}_{m n}+b_{m n} \mathbf{M}_{m n}\right).
        \label{eqn:3}
    \end{equation}
}

\newcommand{\EqnFour}{
    \begin{equation}
        \begin{aligned}
    \mathbf{M}_{m n}=&\left[i \pi_{m n}(\theta) \mathbf{e}_{\theta}-\tau_{m n}(\theta) \mathbf{e}_{\phi}\right] \frac{z_{n}(k r)}{k r} \exp (i m \phi),\\
    \mathbf{N}_{m n}=&\left[\tau_{m n}(\theta) \mathbf{e}_{\theta}+i \pi_{m n}(\theta) \mathbf{e}_{\phi}\right] \frac{z_{n}^{\prime}(k r)}{k r} \mathrm{e}^{i m \phi} \\
    &+\mathbf{e}_{r} n(n+1) P_{n}^{m}(\cos \theta) \frac{z_{n}(k r)}{(k r)^{2}} \mathrm{e}^{i m \phi},
        \end{aligned}
        \label{eqn:4}
    \end{equation}
}

\newcommand{\EqnFive}{
    \begin{equation}
    a_{m n}=\frac{\int_{0}^{2 \pi} \int_{0}^{\pi} \mathbf{E}_{\mathrm{s}} \cdot \mathbf{N}_{m n}^{*} \sin \theta d \theta d \varphi}{i E_{m n} \int_{0}^{2 \pi} \int_{0}^{\pi}\left|\mathbf{N}_{m n}\right|^{2} \sin \theta d \theta d \varphi}.
        \label{eqn:5}
    \end{equation}
}

\newcommand{\EqnSix}{
    \begin{equation}
    \begin{aligned}
    C_{\mathrm{sca}}=\frac{4 \pi}{k^{2}} \sum_{n=1}^{\infty} \sum_{m=-n}^{n} n(n+1)(2 n+1) \\
    \times\frac{(n-m) !}{(n+m) !}\left(\left|\bar{a}_{m n}\right|^{2}+\left|\bar{b}_{m n}\right|^{2}\right),
    \end{aligned}
    \label{eqn:6}
    \end{equation}
}

\newcommand{\EqnSeven}{
    \begin{equation}
        \begin{array}{l}
            p_{m n}^{0}=\frac{1}{n(n+1)}\left[\tau_{m n}(\alpha) \cos \beta-i \pi_{m n}(\alpha) \sin \beta\right], \\
            q_{m n}^{0}=\frac{1}{n(n+1)}\left[\pi_{m n}(\alpha) \cos \beta-i \tau_{m n}(\alpha) \sin \beta\right].
        \end{array}
    \label{eqn:7}
    \end{equation}
}

\newcommand{\EqnEight}{
    \begin{equation}
        \begin{aligned}
         &a_{n}=\frac{\mu \gamma^{2} j_{n}(\gamma x)\left[x j_{n}(x)\right]^{\prime}-\mu_{1} j_{n}(x)\left[\gamma x j_{n}(\gamma x)\right]^{\prime}}{\mu \gamma^{2} j_{n}(\gamma x)\left[x h_{n}^{(1)}(x)\right]^{\prime}-\mu_{1} h_{n}^{(1)}(x)\left[\gamma x j_{n}(\gamma x)\right]^{\prime}}, \\
        &b_{n}=\frac{\mu_{1} j_{n}(\gamma x)\left[x j_{n}(x)\right]^{\prime}-\mu j_{n}(x)\left[\gamma x j_{n}(\gamma x)\right]^{\prime}}{\mu_{1} j_{n}(\gamma x)\left[x h_{n}^{(1)}(x)\right]^{\prime}-\mu h_{n}^{(1)}(x)\left[\gamma x j_{n}(\gamma x)\right]^{\prime}}.
        \end{aligned}
    \label{eqn:8}
    \end{equation}
}

\newcommand{\EqnNine}{
    \begin{equation}
    a_{n}=-\frac{j_{n}(x)\left[x h_{n}^{(1)}(x)\right]^{\prime}}{\left[h_{n}^{(1)}(x)\right]^{\prime}}, 
        b_{n}=-\frac{\left[x j_{n}(x)\right]^{\prime}}{\left[x h_{n}^{(1)}(x)\right]^{\prime}},
    \label{eqn:9}
    \end{equation}
}

\newcommand{\EqnTen}{
    \begin{equation}
    C_{\mathrm{sca}}=\frac{2 \pi}{k^{2}} \sum_{n=1}^{\infty}(2 n+1)\left(\left|a_{n}\right|^{2}+\left|b_{n}\right|^{2}\right).
    \label{eqn:10}
    \end{equation}
}
\newcommand{\FigOne}{
    \begin{figure}[t!]
        \centering
        \includegraphics[width=0.8\linewidth]{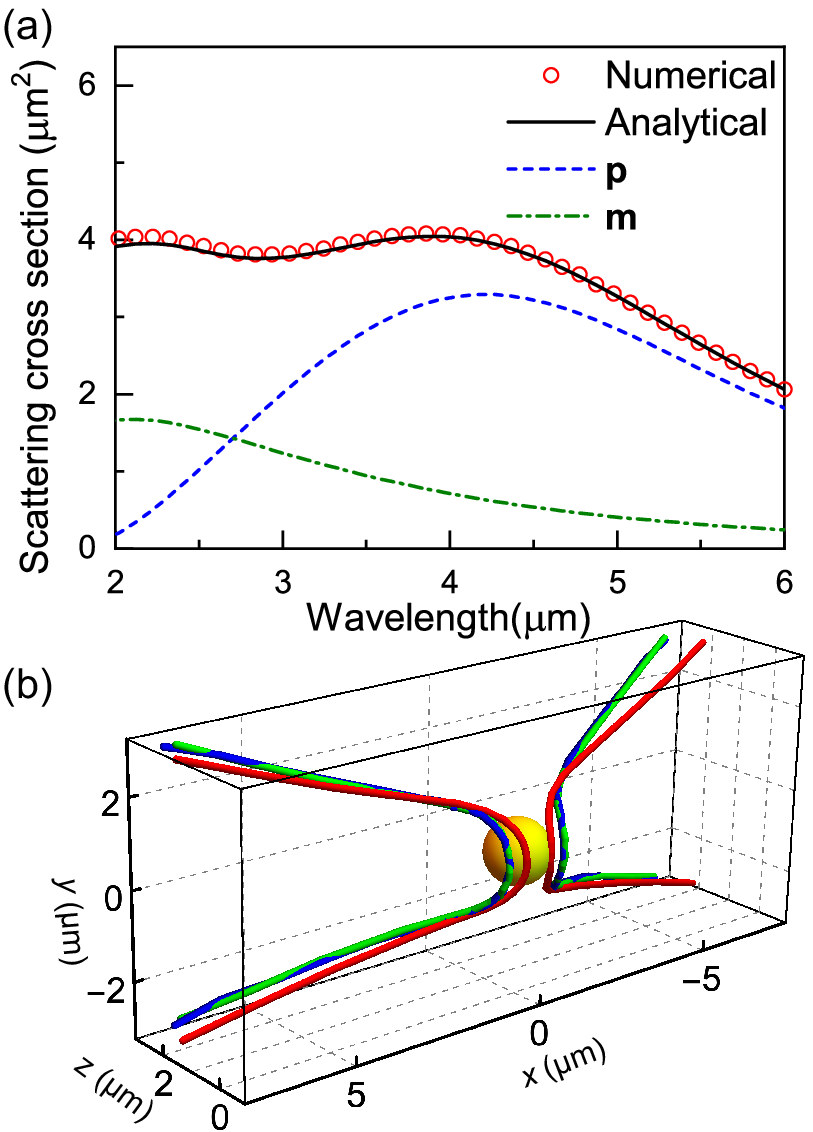}
        \caption{(a) Scattering cross section of the PEC sphere. (b) C lines of electric field for the PEC sphere. The blue solid line denotes the numerical result. The red and green lines denote the analytical results with multipole expansions up to $n=1$ and $n=3$, respectively. The incident wave is linearly polarized along $x$ axis and propagating along $z$ axis.}
        \label{fig:1}
    \end{figure}
}

\newcommand{\FigTwo}{
    \begin{figure*}[t!]
        \centering
        \includegraphics[width=\linewidth]{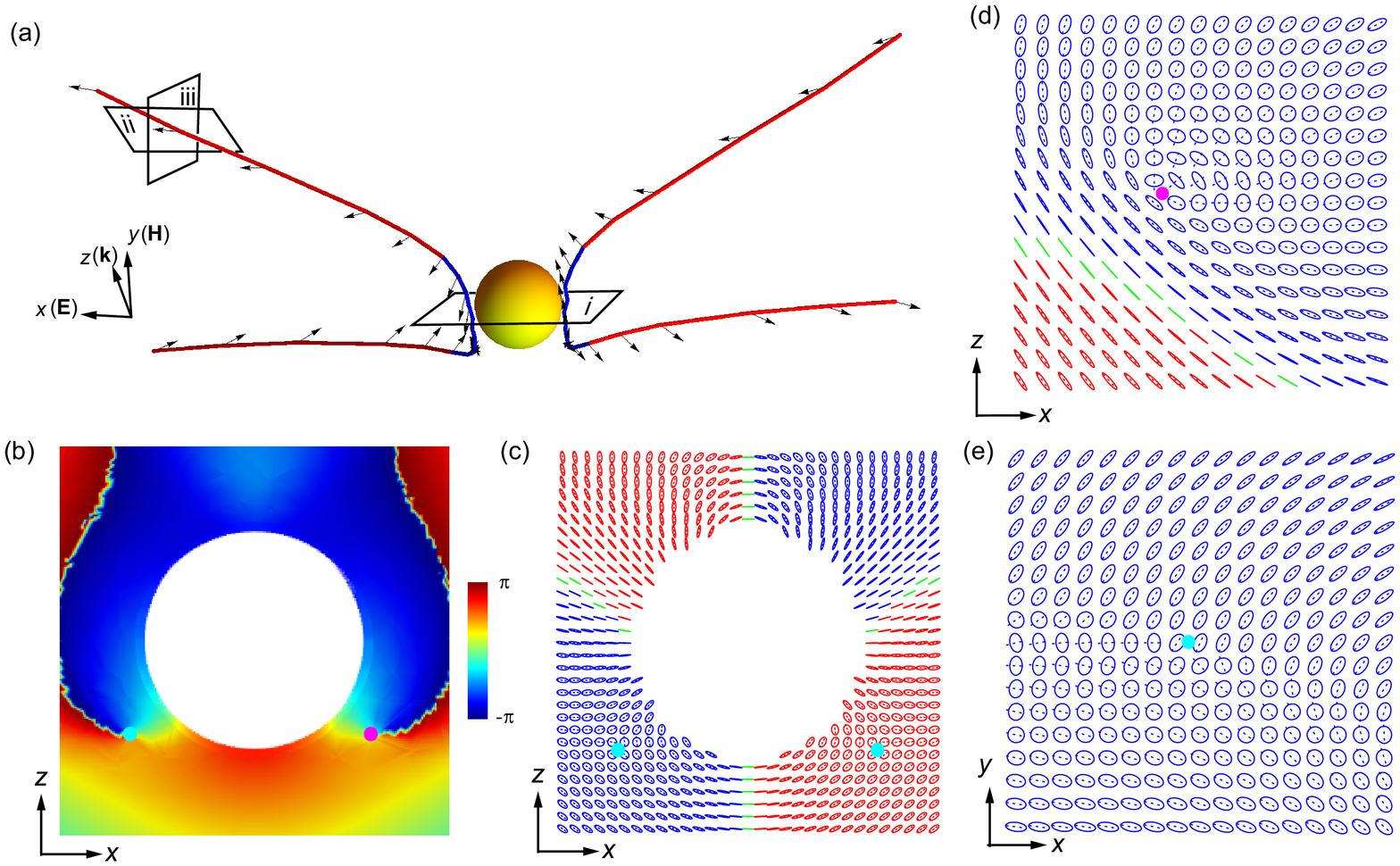}
        \vspace{-100pt}
        \caption{(a) C lines decorated with normal vector $\mathbf{N}$ (black arrows). The C lines are colored in blue (red) to denote a polarization topological index of -1/2 (+1/2). The cutting plane marked as ``i" is a symmetry plane of the system. The planes ``ii" ($xz$-plane) and ``iii" ($xy$-plane) cut the C line at the same point. (b) Phase distribution of $\mathbf{E}\cdot \mathbf{E}$ showing two C points near the sphere on the i-plane. The white region denotes the PEC sphere. (c) Distribution of polarization ellipse on the i-plane. The dashed lines denote the major axis of the polarization ellipses. The blue, red and green polarization ellipses are LCP, RCP and linearly polarized. Polarization ellipses on the (d) ii-plane and (e) iii-plane.}
        \label{fig:2}
    \end{figure*}
}

\newcommand{\FigThree}{
    \begin{figure}[t!]
        \centering
        \includegraphics[width=0.8\linewidth]{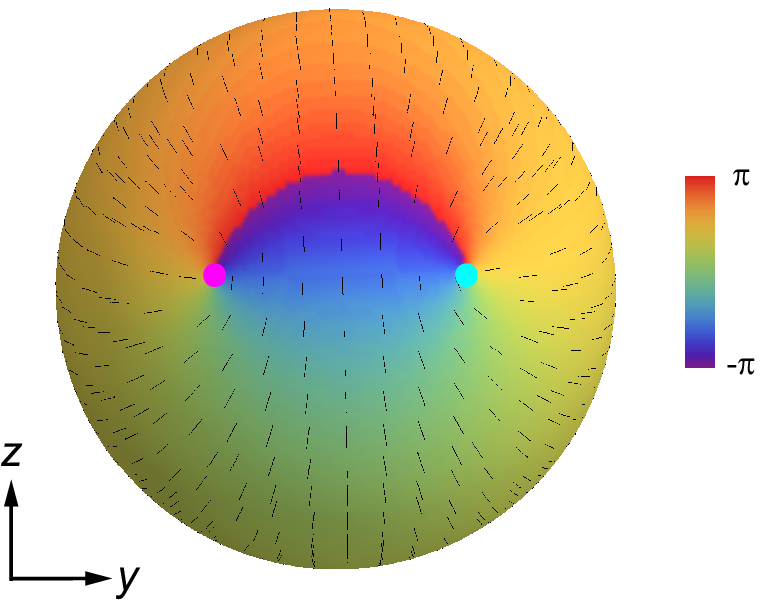}
        \caption{The C points in the far-field region. The color denotes the phase of $\mathbf{E}\cdot \mathbf{E}$, showing the two C points with phase topological indices of $+1$ and $-1$. The line segments denote the major axis of polarization ellipse, showing the two C points are of lemon type and carry the same polarization topological index of $+1/2$. The incident wave is linearly polarized along $x$ axis and propagating along $z$ axis.}
        \label{fig:3}
    \end{figure}
}

\newcommand{\FigFour}{
    \begin{figure}[tb!]
        \centering
        \includegraphics[width=0.8\linewidth]{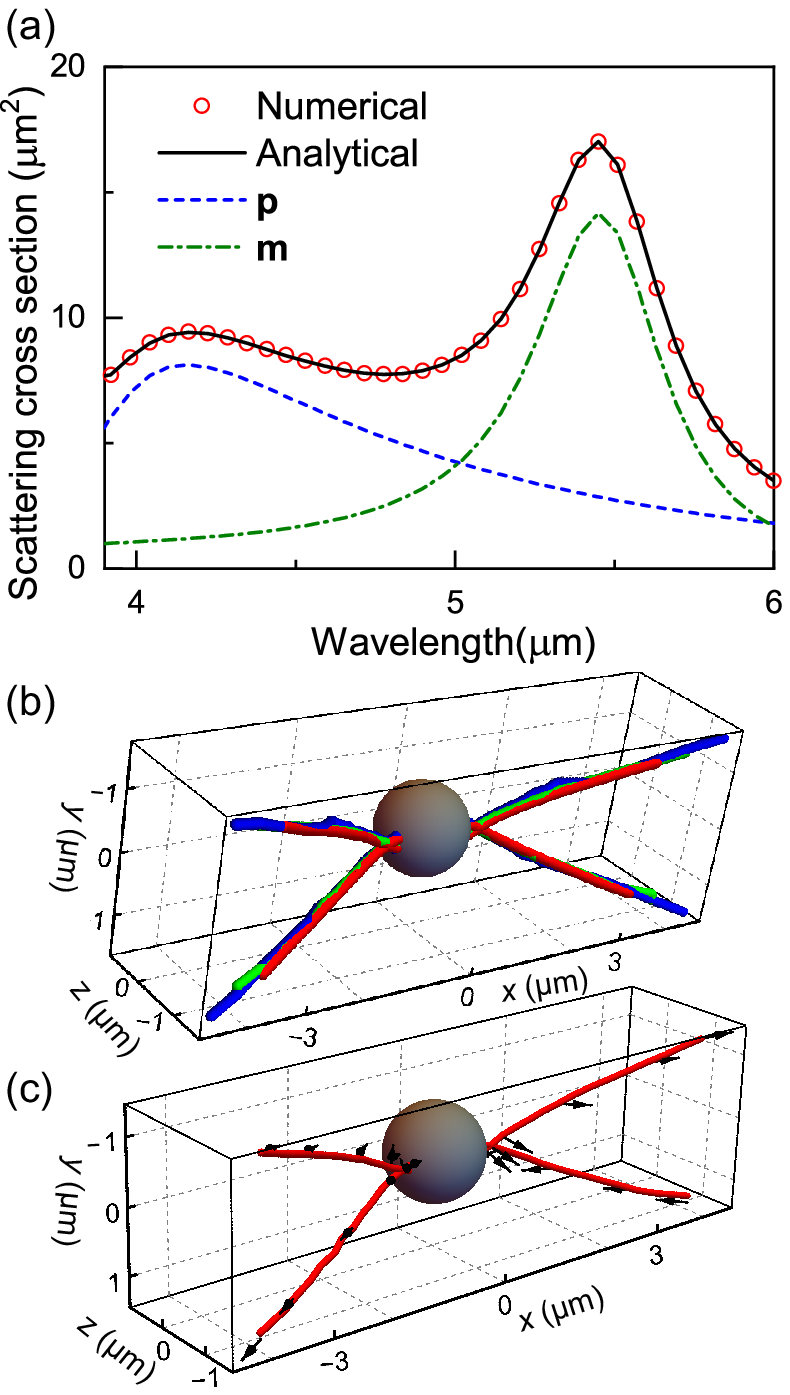}
        \caption{(a) Scattering cross section of the dielectric sphere with $\epsilon_r=12$ under the incidence of a $x$-polarized plane wave propagating in $z$ direction. (b) C lines of electric field for the dielectric sphere. The blue line denotes the numerical result. The red and green lines denote the analytical results with multipole expansions up to $n=1$ and $n=3$, respectively. (c) Normal vector (black arrows) of polarization ellipse on the C lines.}
        \label{fig:4}
    \end{figure}
}

\newcommand{\FigFive}{
    \begin{figure}[tb!]
        \centering
        \includegraphics[width=0.8\linewidth]{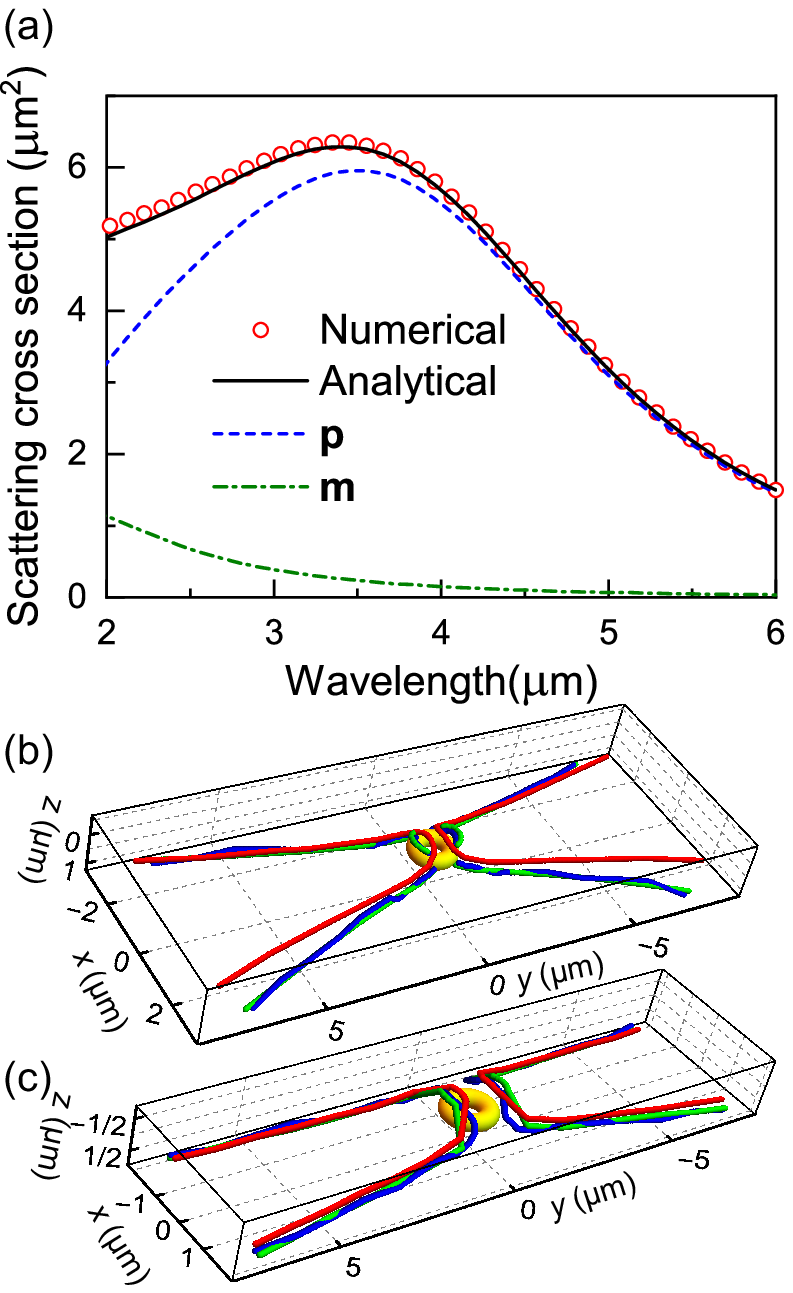}
        \caption{(a) Scattering cross section of the PEC torus under the incidence of a $y$-polarized plane wave propagating in $z$ direction. (b) C lines of electric field for the torus at $\lambda = 2.8$ $\mu$m. (c) C lines of electric field for the torus at $\lambda = 3.6$ $\mu$m. The blue solid line denotes the numerical result. The red and green lines denote the analytical results with multipole expansions up to $n=1$ and $n=3$, respectively.}
        \label{fig:5}
    \end{figure}
}

\newcommand{\FigSix}{
    \begin{figure}[tb!]
        \centering
        \includegraphics[width=0.8\linewidth]{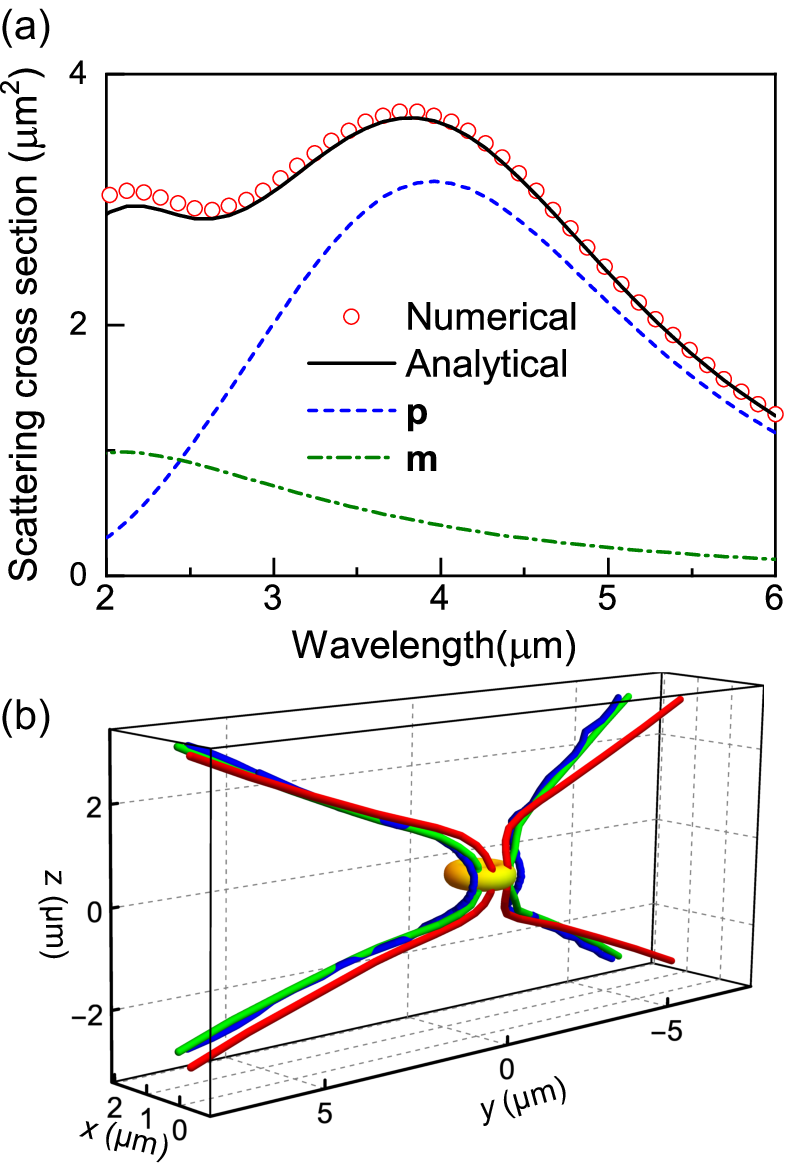}
        \caption{(a) Scattering cross section of the PEC torus under the incidence of a $y$-polarized plane wave propagating in $x$ direction. (b) C lines of electric field for the torus at $\lambda = 3.6$ $\mu$m. The blue solid line denotes the numerical result. The red and green lines denote the analytical results with multipole expansions up to $n=1$ and $n=3$, respectively.}
        \label{fig:6}
    \end{figure}
}

\newcommand{\FigSeven}{
    \begin{figure}[tb!]
        \centering
        \includegraphics[width=0.8\linewidth]{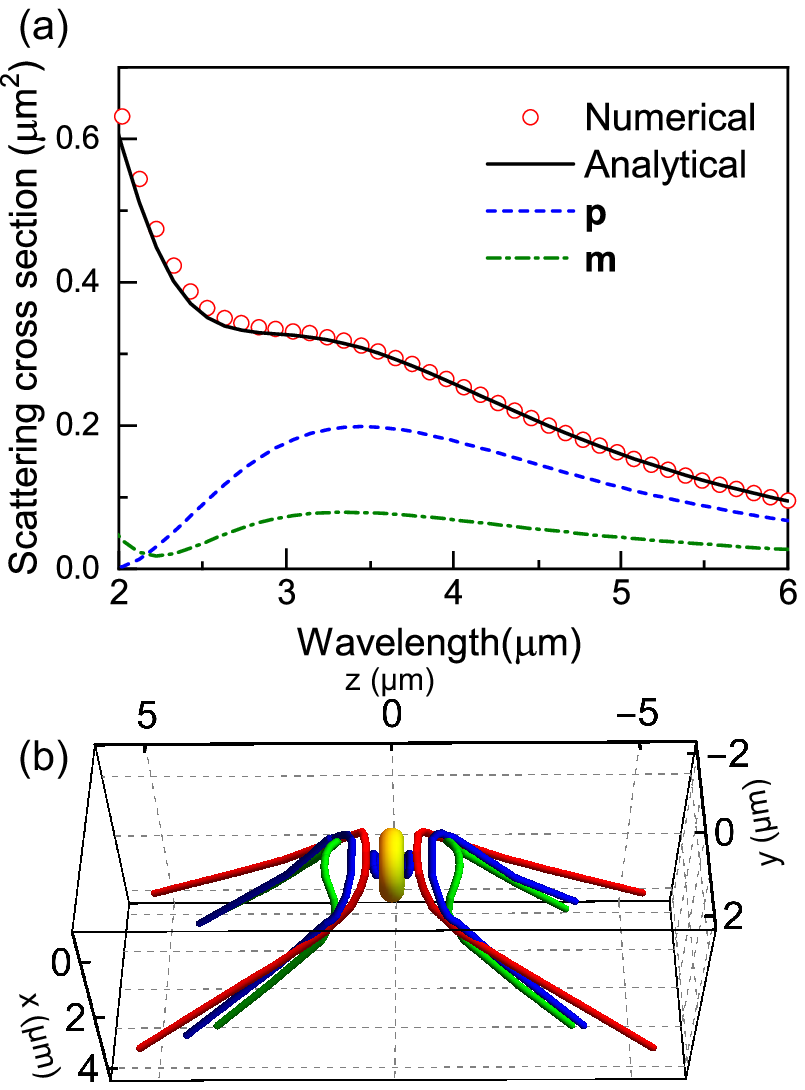}
        \caption{(a) Scattering cross section of the PEC torus under the incidence of a $z$-polarized plane wave propagating in $x$ direction. (b) C lines of electric field for the torus at $\lambda = 3.6$ $\mu$m. The blue solid line denotes the numerical result. The red and green lines denote the analytical results with multipole expansions up to $n=1$ and $n=3$, respectively.}
        \label{fig:7}
    \end{figure}
}

\newcommand{\FigEight}{
    \begin{figure}[tb!]
        \centering
        \includegraphics[width=0.8\linewidth]{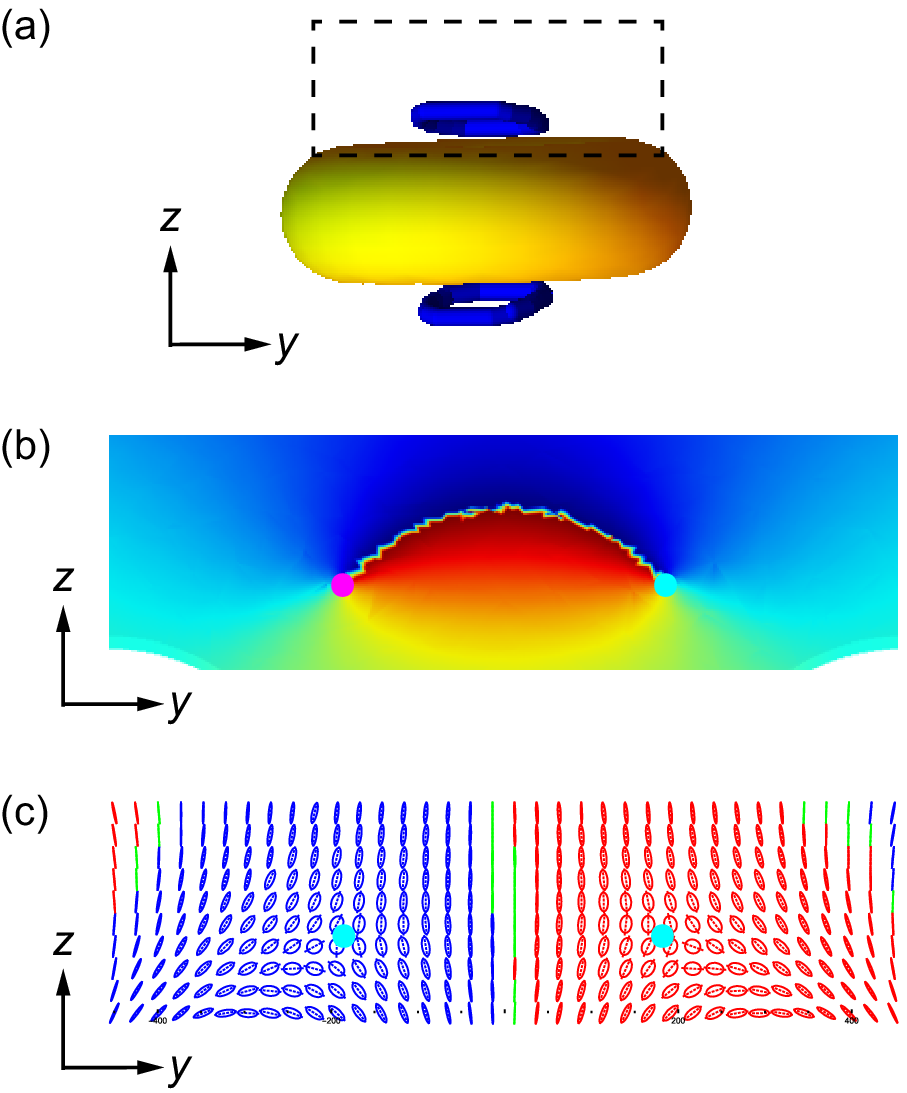}
        \caption{(a) A zoom-in of the C rings. (b) The phase of $\mathbf{E}\cdot \mathbf{E}$ field and (c) the polarization ellipses on the cutting plane marked by the dashed line in (a). The two C points have opposite phase topological indices $\pm 1$ but the same polarization index $-1/2$. Both C points are of star type.}
        \label{fig:8}
    \end{figure}
}

\begin{document}

\title{Polarization singularities in light scattering by small particles}
\date{\today}

\author{Jie Peng}
\affiliation{Department of Physics, City University of Hong Kong, Hong Kong, China}
\author{Wei Liu}
\affiliation{College for Advanced Interdisciplinary Studies, National University of Defense Technology, Changsha, Hunan 410073, China}
\author{Shubo Wang}\email{shubwang@cityu.edu.hk}
\affiliation{Department of Physics, City University of Hong Kong, Hong Kong, China}

\begin{abstract}
Using full-wave numerical simulations and analytical multipole expansions we investigated the properties of real-space polarization singularities that emerge in light scattering by subwavelength particles. We considered spherical and torus particles under the excitation of a linearly polarized plane wave. We determined the topological indices and the trajectories of electric-field polarization singularities in both the near-field and far-field regions. In the far-field region, a total of four singularities are identified and the sum of their polarization topological indices is two, independent of the particle's geometric shape. In the near-field region, the polarization singularities strongly depend on the particle's shape and the polarization of incident light, and their index sum is not governed by the Poincar\'e-Hopf theorem anymore due to the non-transverse nature of the fields. From near field to far field, a flipping of sign can happen to the polarization topological indices of the C lines. The far-field properties of the singularities can be well explained by the interference of the excited multipoles, but their near-field properties can be strongly affected by the evanescent fields that are not captured by the multipole expansions. Our work uncovers the important relationship between particles' geometric properties and the polarization singularities of their scattering field. The results can be applied to manipulate polarization singularities in nanophotonic systems and could generate novel applications in optical sensing.
\end{abstract}
\maketitle

\section{\label{sec: I. Introduction}Introduction}
Polarization singularities are topological defects arising from the spatial distribution of light’s polarization ellipse. It was first introduced by Nye and Hajnal \cite{noauthor_wave_1987} and has received growing attention recently \cite{guo_topologically_2017,larocque_reconstructing_2018,sugic_singular_2018,ariyawansa_polarization_2019,pisanty_knotting_2019,kotlyar_energy_2019,hancock_free-space_2019,bliokh_geometric_2019,wozniak_polarization_2019,chen_singularities_2019}. There are in general two types of polarization singularities: C point and L point. At a C point, light is purely circularly polarized, and the orientation (semi-major axis) of the polarization ellipse is not well defined. At an L point, light is purely linearly polarized, and normal vector of polarization ellipse is not well defined. Both C points and L points can easily emerge in structured light fields. In 2D (two-dimensional) configurations, C points are usually isolated points on the 2D surface, and L points usually form lines (i.e. L lines). In 3D (three-dimensional) configurations, C points normally form C lines, and L lines form L surfaces which split the space into two regions with opposite handedness of elliptical polarization. Recently it was found that polarization singularities have subtle connections with many important concepts such as BIC (bound states in the continuum) \cite{zhen_topological_2014,guo_topologically_2017,zhang_observation_2018,liu_circularly_2019}, electromagnetic multipoles \cite{PhysRevLett.122.153907,doi:10.1002/lpor.202000049,chen_global_2020}, and exceptional points in non-Hermitian systems \cite{noauthor_observation_nodate}. The intriguing properties of polarization singularities can give rise to novel applications such as topologically protected polarization conversion \cite{guo_topologically_2017} and direct visualization of Chern numbers of topological condensed matter systems \cite{fosel_l_2017}. 

Polarization singularities have been theoretically studied in various systems including electric and magnetic fields of paraxial waves \cite{dennis_polarization_2002,berry_electric_2004}, Gaussian beams \cite{freund_polarization_2002} and general 3D fields \cite{berry_index_2004}. Generation and observation of polarization singularities have also been achieved experimentally \cite{angelsky_interferometric_2002,flossmann_polarization_2005,flossmann_polarization_2008,burresi_observation_2009,kurzynowski_regular_2010,park_polarization_2010,vyas_polarization_2013,bauer_observation_2015,ye_simulation_2017,kotlyar_single_2019,chen_observing_2019}. These studies have uncovered many interesting topological properties of polarization singularities \cite{berry_index_2004,kumar_pal_c-point_2017,pal_polarization_2017,chen_singularities_2019}. Based on their winding numbers, C points can be categorized into three generic types, i.e., “star”, “monstar”, and “lemon”, the topological indices of which are respectively -1/2, 1/2, and 1/2. In addition to the various types of polarization patterns, polarization singularities can induce intriguing topological structures in real space. It was shown that the polarization ellipse along a closed path around a C line can form a M\"{o}bius strip \cite{garcia-etxarri_optical_2017}, and C lines can form various types of knots \cite{larocque_reconstructing_2018,PhysRevA.98.063805,sugic_singular_2018,pisanty_knotting_2019,YuKuznetsov:20}. The nontrivial topological properties of C points are broadly connected with many important concepts including geometric phases \cite{berry_geometry_n.d.,quantal_phase_n.d.,bliokh_geometric_2019} and optical orbital angular momentum \cite{kotlyar_energy_2019,hancock_free-space_2019,shen_optical_2019}.

Polarization singularities can emerge in photonic structures with light-matter interactions and their associated properties have recently drawn considerable attention \cite{bauer_observation_2015,garcia-etxarri_optical_2017,liu_circularly_2019,ye_singular_2020}. In this paper, we investigate the properties of C points in simple scenarios of light scattering by subwavelength particles. We determined the C lines in both the far-field and near-field regions via full-wave simulations. It was found that the C lines have similar properties in the far-field in spite of different geometric shapes of the particles and polarizations of incident light. Interestingly, in the near-field region, the structure of C lines strongly depends on the geometric and material properties of the particles. To understand the physics behind the C lines, we applied multipole expansion method to decompose the scattering electric field. We found that the far-field properties of C lines are attributed to the electric and magnetic dipoles, but their near-field properties can be significantly altered by higher-order multipoles. The C lines re-constructed from the multipoles’ fields agree with full-wave numerical results in the far-field region, but significant deviation can appear in the near-field region due to the retardation effect. By comparing the results of a PEC (perfect-electric-conductor) sphere and that of a PEC torus, we show that the geometric topology of scatterers can affect the near-field C lines and give rise to novel structures like nodal rings. In addition, variation of the polarization of incident light can dramatically change the near-field properties of C lines.

The paper is organized as follows. In Sec.\ \ref{sec: II. PS: formulations and visualization}, we briefly introduce the characterization of polarization singularities and the analytical method used in our study. In Sec.\ \ref{sec: III. PS in light scattering by a spherical particle}, we show the results of C lines in the scattering electric field of a PEC sphere and a dielectric sphere. In Sec.\ \ref{sec: IV.  Polarization singularities in light scattering by a torus particle}, we show the results of C lines in the scattering electric field of a metallic torus particle. In both cases, we compare our numerical results with the analytical results based on multipole expansion method. We then draw conclusion in Sec.\ \ref{sec: V.  Conclusion}.

\section{\label{sec: II. PS: formulations and visualization}Polarization singularities: formulations and visualization}
A general 3D electromagnetic field can be characterized by a polarization ellipse with the major axis $\mathbf{A}$ and the minor axis $\mathbf{B}$: $\mathbf{E}=(\mathbf{A}+i\mathbf{B})\operatorname{exp}(i\theta)$. Here, $\mathbf{A}$ and $\mathbf{B}$ satisfy $\mathbf{A}\cdot \mathbf{B}=0$, ${|\mathbf{A}|}\geq {|\mathbf{B}|}$,  and $\theta$ is a proper phase constant. For electric field, we have \cite{dennis_polarization_2002,berry_index_2004}:
\EqnOne
The normal vector of the polarization ellipse can be defined as $\mathbf{N}=2\mathbf{A}\times \mathbf{B}=\operatorname{Im}[\mathbf{E}^*\times \mathbf{E}]$, which indicates the rotation direction of the elliptically polarized electric field (i.e. direction of the electric spin). 

A straightforward way to visualize the C points of electric field is to plot the distribution of polarization ellipses in space and identifying the singularities with well-defined winding number. However, this may not be possible in general 3D electric field whose polarization ellipses are not confined on a 2D plane \cite{bliokh_geometric_2019}. Another way of visualization is to plot the phase of the scalar field $\Psi=\mathbf{E} \cdot \mathbf{E}=E^2 \operatorname{Exp}(i\varphi)$, which is ill-defined at the C points. The topological properties of the C points can be characterized by both the polarization ellipse and the phase $\varphi$, and, correspondingly, two independent topological indices can be assigned to the same C point: the polarization topological index $I_\mathrm{pl}$ and phase topological index $I_\mathrm{ph}$. They can be determined as
\EqnTwo
Here $\phi$ denotes the azimuthal angle of the polarization state on the Poincaré sphere, and both integrals are defined on a closed contour around the C point locating at the origin of $xoy$-plane in which $\mathbf{E}$ field is confined. The indices are positive if $\phi$ or $\varphi$ grows in counterclockwise direction when looking in the $-\hat{z}$ direction. Physically, $I_\mathrm{pl}$ indicates the number of turns that the major axis of polarization ellipse rotates around the C point, and $I_\mathrm{ph}/2$ indicates the number of turns that the electric field vector rotates around the normal vector $\mathbf{N}$ of the polarization ellipse.

We consider subwavelength particles under the excitation of an electromagnetic plane wave. To determine the C points, we first conduct full-wave simulations using COMSOL Multiphysics \cite{COMSOL} to obtain the scattering electric field. The positions of C points are then determined by evaluating the phase $\varphi$. The C lines formed by C points are tracked from the near-field region to the far-field region. To understand how the C points emerge in the systems, we apply multipole expansions to decompose the scattering field into partial waves as  \cite{Rose_Multipole_1955,xu_electromagnetic_n.d.}
\EqnThree
Here $E_{m n}=E_0 i^n (2n+1)(n-m)!/(n+m)!$ , $a_{m n}$ and $b_{m n}$ are the scattering coefficients, $\mathbf{N}_{m n}$ and $\mathbf{M}_{m n}$  are the vector spherical harmonics defined as:
\EqnFour
where $P_n^m(x)$ is the associated Legendre function of the first kind, $\pi_{m n}(\theta)=m/\sin\theta P_n^m(\cos\theta)$ and $\tau_{m n}(\theta)=\frac{d}{d\theta}P_n^m(\cos\theta)$, and $z_n (x)=x h_n^{(1)}(x)$ with $h_n^{(1)}(x)$ being the first kind of spherical Hankel function. 

For particles of arbitrary shape, the scattering coefficients in Eq.\ (\ref{eqn:3}) can be determined as, taking $a_{m n}$ for instance,
\EqnFive
Under plane wave incidence, the normalized scattering cross section can be determined as:
\EqnSix
where $\bar{a}_{m n}=p_{m n}^0 a_{m n}$, $\bar{b}_{m n}=q_{m n}^0 b_{m n}$ and
\EqnSeven
Here, $p_{m n}^0$  and $q_{m n}^0$ are the expansion coefficients of the incident wave, $\alpha$ and $\beta$ are the polar angle and azimuthal angle of the incident wavevector $\mathbf{k}$, respectively.
For dielectric spheres under the incidence of linearly polarized plane wave, the scattering coefficients (for which $m=1$) correspond to the Mie coefficients \cite{bohren_absorption_2004}:
\EqnEight
Here $x=k a=2\pi Na/\lambda$ is the size parameter and $\gamma=k_1/k=N_1/N$ is the relative refractive index. $N_1$ and $N$ are the refractive indices of the particle and background medium, respectively. $\mu_1$  and $\mu$ are relative permeability of the sphere and background medium, respectively. $j_n(x)$ is the spherical Bessel function. 

For spherical particle made of PEC, the Mie coefficients are reduced to  \cite{ruppin_scattering_2006}:
\EqnNine
With these coefficients, the scattering cross section can be rewritten as:
\EqnTen
The total scattering cross section can also be evaluated directly via integrating the Poynting vector of the scattering field over a surface enclosing the particle, which should agree with the result of Eqs.\ (\ref{eqn:6}) or (\ref{eqn:10}) when $n \rightarrow \infty$. Equations (\ref{eqn:6}) and (\ref{eqn:10}) can be applied to analyze the dominating components of the multipoles. To understand the contribution of individual multipoles to the C lines, we apply Eq.\ (\ref{eqn:3}) to reconstruct the electric field and C lines with $n$ truncated to a finite value.

\section{\label{sec: III. PS in light scattering by a spherical particle}Polarization singularities in light scattering by a spherical particle}
We first consider a spherical PEC particle of radius $r = 750$ nm under the incidence of a plane wave linearly polarized along $x$ axis and propagating along $z$ axis. To characterize the scattering properties of the particle in the subwavelength regime, we numerically calculated its scattering cross section within the wavelength range of [2 $\mu$m, 6 $\mu$m]. The results are shown in Fig.\ \ref{fig:1}(a) as the circle-symbol line. Since the PEC material does not support plasmonic resonance, the scattering cross section does not have any evident peak. However, we can still identify a local maximum at the wavelength $\lambda=4.0$ $\mu$m. The C lines at this wavelength is determined numerically by plotting $\varphi$, and the result is denoted by the blue line in Fig.\ \ref{fig:1}(b). We notice that a total of four C lines are identified in the far-field region, which are approximately straight lines extending to infinity. In the near-field region, the four C lines merge in pairs. Alternatively, one can consider that there are two C lines extending from the far field to the near field and back to the far field.

\FigOne
\FigTwo
To understand why the C lines emerge in the scattering field, we apply multipole expansions as detailed in Sec. II to decompose the electric field and study the contribution of multipoles to the scattering cross section and C lines. The solid black line in Fig.\ \ref{fig:1}(a) shows the scattering cross section calculated using Eq.\ (\ref{eqn:10}) by truncating the order to $n = 3$, which agrees with the full-wave numerical result. The contributions of electric dipole $\mathbf{p}$ and magnetic dipole $\mathbf{m}$ are denoted by the blue and green dashed lines, respectively. Evidently, the considered local maximum is mainly attributed to the electric dipole resonance. Using Eq.\ (\ref{eqn:3}), we then analytically evaluated the scattering field and re-construct the C lines at $\lambda = 4.0$ $\mu$m. The results are denoted as the red and green lines in Fig\ \ref{fig:1}(b) for the truncation order of $n = 1$ and $n = 3$, respectively. We notice that the analytical C lines with $n = 1$ can already reproduce the essential features of the C lines (e.g., the total number of lines and their trend). Thus, we can conclude that the interference of $\mathbf{p}$ and $\mathbf{m}$ is responsible for the emergence of C lines. As expected, the analytical C lines with $n = 3$ well agree with the numerical result. 

The topological properties of the C lines for the PEC sphere can be studied by plotting the phase $\varphi$ and the polarization ellipses in space. Since the scattering electric field in the near-field region is not confined on a 2D plane, the corresponding polarization index is generally ill-defined. However, on a symmetry plane such as the $xoz$-plane marked by the black rectangle ``i" in Fig. 2(a), the electric field is purely tangent to the plane, thus, both the polarization topological index and phase topological index can be defined. Figure \ref{fig:2}(b) shows the distribution of phase $\varphi$ on this plane, where two C points appear in the vicinity of the sphere (corresponding to the white region), as marked by the magenta point for $I_\mathrm{ph}=+1$ and cyan point for $I_\mathrm{ph}=-1$. The distribution of polarization ellipses on the same plane is shown in Fig.\ \ref{fig:2}(c), where the blue (red) ellipses correspond to LCP (RCP) polarization with $\mathbf{A}\times \mathbf{B}<0$ ( $\mathbf{A}\times \mathbf{B}>0$) and the green line segments correspond to linear polarization. The dashed lines inside the ellipses denote the polarization major axis. We notice that the polarization ellipses near the two C points have opposite chirality corresponding to opposite spin. In addition, both C points are of star type and carry the same polarization topological index $I_\mathrm{pl}=-1/2$, as marked by the cyan dots. Therefore, the polarization topological index of a C point is generally different from its phase topological index \cite{bliokh_geometric_2019}. For the near-field C points away from the symmetry plane, their polarization distribution manifest different topological properties in different cutting planes due to the 3D nature of the electric field. This can be verified by considering two different planes ``ii" ($xz$-plane) and ``iii" ($xy$-plane) as marked in Fig. \ref{fig:2}(a), which cut the C line at the same point. The polarization ellipses on the two planes are shown in Fig. \ref{fig:2}(d) and 2(e), respectively. As expected, the C point manifests different topological properties on the two planes, i.e. monstar ($I_\mathrm{pl}=+1/2$) in Fig.\ \ref{fig:2}(d) and star ($I_\mathrm{pl}=-1/2$) in Fig.\ \ref{fig:2}(e).

\FigThree

For C points in the far-field region, both the phase topological index and the polarization topological index are well defined since the scattering far field is orthogonal to the wavevector and locally approximates a plane wave, i.e. it is a 2D vector field lies on the far-field $\hat{r}$ sphere. Figure \ref{fig:3} shows the distribution of $\varphi$ and $\mathbf{A}$ on one side of the far-field $\hat{r}$ sphere for the PEC sphere. Evidently, two C points of lemon type appear, and another two C points appear symmetrically on the opposite side. Remarkably, the C points carry opposite phase topological indices $I_\mathrm{ph}=\pm 1$ but the same polarization topological indices  $I_\mathrm{pl}=+1/2$, in contrast to the near-field C points which have $I_\mathrm{pl}=-1/2$. The different $I_\mathrm{pl}$ of the far-field and near-field C points indicate a sign change of the index. The transition can be visualized by plotting the normal vector $\mathbf{N}$ along the C lines, as shown in Fig.\ref{fig:2}(a) by the black arrows. We emphasize that the index sign flipping would happen around the point where $\mathbf{N}$ is normal to the C line \cite{noauthor_wave_1987}. This is visualized in Fig. \ref{fig:2}(a) through the colour change of the C lines. In addition, the sum of the topological indices of all the four C points in far-field region is $\sum I_\mathrm{pl} =2$ in accordance with the Poincaré-Hopf theorem \cite{frankel_geometry_2012}. While in the near field, the index sum is not bounded to be 2, simply due to the fact that the fields are not transverse (tangent) anymore.

\FigFour

For comparison, we also consider the C lines in the case of a dielectric sphere. The sphere has the same radius $r = 750$ nm and is made of silicon with relative permittivity $\epsilon_d=12$. Under the excitation of a plane wave propagating along $z$ axis and polarized along $x$ axis, its scattering cross section has a peak at $\lambda = 5.4$ $\mu$m, as shown in Fig.\ \ref{fig:4}(a) by the circle-symbol line.  The C lines at this wavelength is shown in Fig.\ \ref{fig:4}(b) by the solid blue line. We notice that a total of four C lines emerge in the far-field region, similar to the case of the PEC sphere. However, the C lines do not merge in the near-field region. Instead, each pair of the C lines forms a crossing and is connected to the surface of the silicon sphere. This near-field “crossing” feature has not been observed in previous studies, as far as we know. To understand these features of C lines, we apply multipole expansions to the scattering field and analytically evaluated the scattering cross section. The results with truncation order $n = 3$ and the individual contributions of $\mathbf{p}$ and $\mathbf{m}$ are denoted by the solid black line, the dashed blue and green lines in Fig.\ \ref{fig:4}(a), respectively. We notice that the multipoles with order up to $n = 3$ can almost reproduce the scattering cross section, and the contributions of $\mathbf{p}$ and $\mathbf{m}$ dominate in the considered wavelength range. We also notice that the peak at $\lambda = 5.4$ $\mu$m is mainly attributed to the magnetic dipole resonance, in contrast to the PEC case in Fig.\ \ref{fig:1}(a) where the first peak is dominated by the electric dipole resonance. This may explain the different near-field properties of C lines in the PEC and dielectric spheres. We then analytically re-constructed the C lines by truncating the multipoles to $n = 1$ and $n = 3$, and the results are denoted as solid red and green lines in Fig.\ \ref{fig:4}(b). We see that both agree with the numerical result (i.e. blue line), indicating that higher-order multipoles other than $\mathbf{p}$ and $\mathbf{m}$ are negligible. We note that the contributions of both $\mathbf{p}$ and $\mathbf{m}$ are important for the emergence of C lines. The field of a single $\mathbf{p}$ or $\mathbf{m}$ cannot give rise to C lines \cite{garcia-etxarri_optical_2017}. 

In the far-field region, we found that the phase and polarization topological indices of the C lines are the same as that of the PEC sphere. However, the topological indices are dramatically different in the near-field region. In contrast to the PEC sphere where the polarization topological index undergoes a sign change from near field to far field, the index remains the same (i.e. $+1/2$) for all the four C lines in Fig. \ref{fig:4}(b) even after the ``crossing". This is confirmed by plotting the normal vector $\mathbf{N}$ on the C lines, as shown in Fig. \ref{fig:4}(c) by the black arrows. It is noticed that the angle between $\mathbf{N}$ and the C lines is less than $\pi/2$ in both the near-field and far-field regions. The sign flipping of the index requires that $\mathbf{N}$ is normal to the C line, which is not observed in Fig. \ref{fig:4}(c).

 The comparison between the PEC sphere and dielectric sphere shows that the major features of C lines (e.g. total number and trend) are determined by the multipole components, while the material properties can affect the relative weighting of multipoles and the near-field features of C lines. In the following, we will focus on particles made of PEC and investigate the effects of geometric topology and the polarization of incident light.

\FigFive

\section{\label{sec: IV. Polarization singularities in light scattering by a torus particle}Polarization singularities in light scattering by a torus particle}
To understand how the particle’s geometry affect the topological properties of C points, we consider a small torus particle under the incidence of a plane wave propagating along $z$ axis and polarized along $y$ axis, as shown in Fig.\ \ref{fig:5}(b). The torus lies on the $xoy$-plane and has an inner radius of $250$ nm and an outer radius of $750$ nm. In contrast to the spherical particle whose surface has genus (i.e. number of “holes”) $g = 0$, the torus particle has a geometric surface with genus $g = 1$, where the Gaussian curvature can be either positive or negative. The geometric properties of the torus particle can give rise to interesting features of C lines in the near-field region. Figure \ref{fig:5}(a) shows numerical (circle-symbol line) and analytical (solid and dashed lines) results of the scattering cross section of the torus particle. The C lines at the wavelength $\lambda = 2.8$ $\mu$m is shown in Fig.\ \ref{fig:5}(b) as the solid blue line. We notice that the features of the C lines are similar to that of the sphere case. Interestingly, the C lines  have four “handles” connected to the torus surface: two locate above the torus and two locate below the torus. The emergence of the C-line structures can be understood by applying multipole expansions to decompose the scattering field and scattering cross section. As shown in Fig.\ \ref{fig:5}(a), the analytical result of scattering cross section with truncation $n = 3$ (solid black line) well agrees with the numerical result. The scattering cross section are mainly contributed by $\mathbf{p}$ (dashed blue line) and $\mathbf{m}$ (dashed green line). The analytical C lines obtained by multipole expansions with orders up to $n = 1$ and $n = 3$ are shown as red and green lines in Fig.\ \ref{fig:5}(b), respectively. As expected, the C lines with $n = 1$ (i.e. dipoles) can already reproduce the far-field features, but it cannot give the ``handle" structure in the near-field. The C lines with $n = 3$ agree well with the numerical C lines in the far field. However, there is deviation in the near field which exists even if the truncation order $n$ approaches infinity. This is due to the limitation of multipole expansion method in describing the near fields, since it assumes the multipoles locate at the origin and neglects the retardation effect associated with internal structure of the particle. The deviation between numerical and analytical results in the near-field region is more evident at the resonance wavelength $\lambda = 3.6$ $\mu$m, as shown in Fig. \ref{fig:5}(c), where the C lines with $n=1$ (red line) and $n=3$ (green line) do not agree with the numerical results (blue line) in the vicinity of the torus particle.

 \FigSix
 \FigSeven
 
 To understand how the polarization of incident light affects the C lines, we consider the configuration in Fig.\ \ref{fig:6}(b), where the plane wave propagates along $x$ direction and is polarized along $y$ direction. The numerically calculated scattering cross section is shown in Fig.\ \ref{fig:6}(a) as the circle-symbol line, which is similar to the case of Fig. \ref{fig:5}(a). For consistence and comparison, we determined the C lines at the same wavelength $\lambda = 3.6$ $\mu$m, which is shown as the solid blue line in Fig.\ \ref{fig:6}(b). We notice that the C lines have similar structures as that in Fig.\ \ref{fig:1}(b) and Fig.\ \ref{fig:5}(b). By multipole expansions, we analytically calculated the scattering cross section, and the results are shown in Fig.\ \ref{fig:6}(a) as the solid black line, the dashed blue and green lines for the total scattering cross section with truncation order $n=3$, the individual contributions of $\mathbf{p}$ and $\mathbf{m}$, respectively. Again, the electric dipole dominates at the considered wavelength, and the analytical result with truncation order $n = 3$ well agree with the numerical result. We also analytically construct the C lines for $n = 1$ and $n = 3$ which are shown as the red and green lines in Fig. \ref{fig:6}(b), respectively. The C lines with larger truncation order of $n = 3$ gives better agreement. It can reproduce the features of C lines in the far-field region, but in the near-field region they deviate from the numerical C lines (blue lines) due to the inaccurate description of multipole expansions for the near fields.

The results in Fig. \ref{fig:5}(b) and Fig. \ref{fig:6}(b) show that different polarization of the incident wave can induce different C-line structures in the near-field. To further understand this property, we consider the configuration shown in Fig.\ \ref{fig:7}(b), where the same torus particle is under the excitation of a plane wave polarized along $z$ direction and propagating along $x$ direction. The numerically obtained scattering cross section for this case is denoted as the circle-symbol line in Fig. \ref{fig:7}(a). Notably, the scattering cross section is one order of magnitude smaller than the cases of Fig. \ref{fig:5}(a) and Fig. \ref{fig:6}(a). The numerically determined C lines at $\lambda = 3.6$ $\mu$m are shown as the solid blue line in Fig.\ \ref{fig:7}(b). It is noticed that the C lines have similar far-field features as in the case of Fig.\ \ref{fig:5}(b) and (c). Interestingly, a pair of C rings emerges in the near-field region: one is on the left side and the other is the on the right side. We will discuss the topological properties of the C rings later. By applying multipole expansions, we analytically evaluate the scattering cross section of the torus particle. The results for truncation order up to $n = 3$ and the individual contributions of $\mathbf{p}$ and $\mathbf{m}$ are shown as the solid black line, the dashed blue and green lines in Fig.\ \ref{fig:7}(a), respectively. In contrast to previous cases, the contributions of $\mathbf{p}$ and $\mathbf{m}$ are comparable, which may explain the emergence of the C rings. The analytical C lines with truncation order $n = 1$ and $n = 3$ are denoted as the red and green lines in Fig.\ \ref{fig:7}(b). It is noticed that increasing the truncation order from $n = 1$ to $n = 3$ leads to better agreement between the analytical and numerical C lines in the far-field region. However, the near-field C rings cannot be reproduced by the analytical method of multipole expansions due to the neglect of retardation effect.

\FigEight
 
The C rings emerging in the vicinity of the torus can have interesting topological properties. A zoom-in of the rings is shown in Fig.\ \ref{fig:8}(a). The two rings are symmetric with respect to the $xoy$-plane due to the symmetry of the torus and the incident light. Figure\ \ref{fig:8}(b) shows the distribution of phase $\varphi$ on the $yoz$-plane within the region marked by the dashed rectangle in Fig.\ \ref{fig:8}(a). The C points on this cutting plane are labeled by the cyan and magenta points corresponding to phase topological index of $I_\mathrm{ph}=-1$  and $I_\mathrm{ph}=+1$, respectively. We note that $I_\mathrm{ph}$ of the two C points will carry the same sign if we define the positive direction as the evolving direction of the C line. In addition, we determined the distribution of the polarization ellipses on the same cutting plane, and the result is shown in Fig.\ \ref{fig:8}(c). The polarization topological indices of the C points are well defined since the electric field is confined in the symmetry plane of $yoz$. The two C points carry the same polarization topological index of $I_\mathrm{pl}=-1/2$. We note that the polarization ellipses near the C points have opposite chirality, similar to the case of the PEC sphere in Fig.\ \ref{fig:2}(c)\cite{Berry_2013}.

\section{\label{sec: V. Conclusion}Conclusion}
We investigated the C lines emerged in light scattering by small particles. By considering particles of different shape (i.e. sphere and torus) and incident light with different propagating directions and polarizations, we have uncovered the dependence of the C lines on the particles’ geometric properties and the excitation. In the considered subwavelength regime, a total of four C lines are induced in the far-field region regardless of the geometry of the particle and the polarization of incident light. The sum of the polarization topological indices of all four C lines is bouned to be $2$ in the far field, as is required by the Poincaré-Hopf theorem. In addition, interesting structures of C lines such as “handles” and “rings” can emerge in the near-field region. We applied multipole expansions to understand the mechanism for the emergence of the C lines. We found that the C lines are mainly attributed the interference of the induced electric and magnetic dipoles, which can well describe the key features of the C lines in the far-field region. However, the multipole expansions cannot capture certain near-field features which strongly depend on the geometric details of the particles. This indicates that complex and novel structures of polarization singularities can emerge in various subwavelength nanophotonic systems which worth further study. It is worth to emphasize that the relatively simple structures of C lines in the considered systems are attributed to the dominance of electric and magnetic dipoles in the subwavelength regime. When the sizes of the particles are comparable or larger than wavelength (or the dielectric constant is large enough), higher-order multipoles can be excited, which can give rise to more C lines and more sophisticated morphologies for C line distributions \cite{chen_global_2020}. We also note that material dispersion and loss are not considered in the current study, the introduction of which will not significantly affect the C lines as long as the dominating multipoles remain unchanged.

The interactions between chiral light fields and small particles can give rise to many intriguing phenomena such as photonic spin-orbit interactions and optical force manipulations\cite{wang_lateral_2014,bliokh_spinorbit_2015,wang2019arbitrary}. Our study show that such chiral light-matter interactions can also appear in the scattering of linearly polarized light by achiral particles. The results can be applied to generate C lines by using small particles. Since the properties of C lines are rather sensitive to the geometric shape of the particles and the excitations, the results may also find applications in optical sensing via the polarization singularities of light.

\section{Acknowledgements}
The work described in this paper was supported by grants from the Research Grants Council of the Hong Kong Special Administrative Region, China (Project No. CityU 11306019 and No. HKUST C6013-18G).

\bibliography{references_zotero}

\begin{thebibliography}{53}%
\makeatletter
\providecommand \@ifxundefined [1]{%
 \@ifx{#1\undefined}
}%
\providecommand \@ifnum [1]{%
 \ifnum #1\expandafter \@firstoftwo
 \else \expandafter \@secondoftwo
 \fi
}%
\providecommand \@ifx [1]{%
 \ifx #1\expandafter \@firstoftwo
 \else \expandafter \@secondoftwo
 \fi
}%
\providecommand \natexlab [1]{#1}%
\providecommand \enquote  [1]{``#1''}%
\providecommand \bibnamefont  [1]{#1}%
\providecommand \bibfnamefont [1]{#1}%
\providecommand \citenamefont [1]{#1}%
\providecommand \href@noop [0]{\@secondoftwo}%
\providecommand \href [0]{\begingroup \@sanitize@url \@href}%
\providecommand \@href[1]{\@@startlink{#1}\@@href}%
\providecommand \@@href[1]{\endgroup#1\@@endlink}%
\providecommand \@sanitize@url [0]{\catcode `\\12\catcode `\$12\catcode
  `\&12\catcode `\#12\catcode `\^12\catcode `\_12\catcode `\%12\relax}%
\providecommand \@@startlink[1]{}%
\providecommand \@@endlink[0]{}%
\providecommand \url  [0]{\begingroup\@sanitize@url \@url }%
\providecommand \@url [1]{\endgroup\@href {#1}{\urlprefix }}%
\providecommand \urlprefix  [0]{URL }%
\providecommand \Eprint [0]{\href }%
\providecommand \doibase [0]{https://doi.org/}%
\providecommand \selectlanguage [0]{\@gobble}%
\providecommand \bibinfo  [0]{\@secondoftwo}%
\providecommand \bibfield  [0]{\@secondoftwo}%
\providecommand \translation [1]{[#1]}%
\providecommand \BibitemOpen [0]{}%
\providecommand \bibitemStop [0]{}%
\providecommand \bibitemNoStop [0]{.\EOS\space}%
\providecommand \EOS [0]{\spacefactor3000\relax}%
\providecommand \BibitemShut  [1]{\csname bibitem#1\endcsname}%
\let\auto@bib@innerbib\@empty
\bibitem [{\citenamefont {Nye}\ and\ \citenamefont
  {Hajnal}(1987)}]{noauthor_wave_1987}%
  \BibitemOpen
  \bibfield  {author} {\bibinfo {author} {\bibfnamefont {J.~F.}\ \bibnamefont
  {Nye}}\ and\ \bibinfo {author} {\bibfnamefont {J.~V.}\ \bibnamefont
  {Hajnal}},\ }\href {https://doi.org/10.1098/rspa.1987.0002} {\bibfield
  {journal} {\bibinfo  {journal} {Proc. R. Soc. Lond. A}\ }\textbf {\bibinfo
  {volume} {409}},\ \bibinfo {pages} {21} (\bibinfo {year} {1987})}\BibitemShut
  {NoStop}%
\bibitem [{\citenamefont {Guo}\ \emph {et~al.}(2017)\citenamefont {Guo},
  \citenamefont {Xiao},\ and\ \citenamefont {Fan}}]{guo_topologically_2017}%
  \BibitemOpen
  \bibfield  {author} {\bibinfo {author} {\bibfnamefont {Y.}~\bibnamefont
  {Guo}}, \bibinfo {author} {\bibfnamefont {M.}~\bibnamefont {Xiao}},\ and\
  \bibinfo {author} {\bibfnamefont {S.}~\bibnamefont {Fan}},\ }\href
  {https://doi.org/10.1103/PhysRevLett.119.167401} {\bibfield  {journal}
  {\bibinfo  {journal} {Phys. Rev. Lett.}\ }\textbf {\bibinfo {volume} {119}},\
  \bibinfo {pages} {167401} (\bibinfo {year} {2017})}\BibitemShut {NoStop}%
\bibitem [{\citenamefont {Larocque}\ \emph {et~al.}(2018)\citenamefont
  {Larocque}, \citenamefont {Sugic}, \citenamefont {Mortimer}, \citenamefont
  {Taylor}, \citenamefont {Fickler}, \citenamefont {Boyd}, \citenamefont
  {Dennis},\ and\ \citenamefont {Karimi}}]{larocque_reconstructing_2018}%
  \BibitemOpen
  \bibfield  {author} {\bibinfo {author} {\bibfnamefont {H.}~\bibnamefont
  {Larocque}}, \bibinfo {author} {\bibfnamefont {D.}~\bibnamefont {Sugic}},
  \bibinfo {author} {\bibfnamefont {D.}~\bibnamefont {Mortimer}}, \bibinfo
  {author} {\bibfnamefont {A.~J.}\ \bibnamefont {Taylor}}, \bibinfo {author}
  {\bibfnamefont {R.}~\bibnamefont {Fickler}}, \bibinfo {author} {\bibfnamefont
  {R.~W.}\ \bibnamefont {Boyd}}, \bibinfo {author} {\bibfnamefont {M.~R.}\
  \bibnamefont {Dennis}},\ and\ \bibinfo {author} {\bibfnamefont
  {E.}~\bibnamefont {Karimi}},\ }\href
  {https://doi.org/10.1038/s41567-018-0229-2} {\bibfield  {journal} {\bibinfo
  {journal} {Nat. Phys.}\ }\textbf {\bibinfo {volume} {14}},\ \bibinfo {pages}
  {1079} (\bibinfo {year} {2018})}\BibitemShut {NoStop}%
\bibitem [{\citenamefont {Sugic}\ and\ \citenamefont
  {Dennis}(2018)}]{sugic_singular_2018}%
  \BibitemOpen
  \bibfield  {author} {\bibinfo {author} {\bibfnamefont {D.}~\bibnamefont
  {Sugic}}\ and\ \bibinfo {author} {\bibfnamefont {M.~R.}\ \bibnamefont
  {Dennis}},\ }\href {https://doi.org/10.1364/JOSAA.35.001987} {\bibfield
  {journal} {\bibinfo  {journal} {J. Opt. Soc. Am. A}\ }\textbf {\bibinfo
  {volume} {35}},\ \bibinfo {pages} {1987} (\bibinfo {year}
  {2018})}\BibitemShut {NoStop}%
\bibitem [{\citenamefont {Ariyawansa}\ \emph {et~al.}(2019)\citenamefont
  {Ariyawansa}, \citenamefont {Liang},\ and\ \citenamefont
  {Brown}}]{ariyawansa_polarization_2019}%
  \BibitemOpen
  \bibfield  {author} {\bibinfo {author} {\bibfnamefont {A.}~\bibnamefont
  {Ariyawansa}}, \bibinfo {author} {\bibfnamefont {K.}~\bibnamefont {Liang}},\
  and\ \bibinfo {author} {\bibfnamefont {T.~G.}\ \bibnamefont {Brown}},\ }\href
  {https://doi.org/10.1364/JOSAA.36.000312} {\bibfield  {journal} {\bibinfo
  {journal} {J. Opt. Soc. Am. A}\ }\textbf {\bibinfo {volume} {36}},\ \bibinfo
  {pages} {312} (\bibinfo {year} {2019})}\BibitemShut {NoStop}%
\bibitem [{\citenamefont {Pisanty}\ \emph {et~al.}(2019)\citenamefont
  {Pisanty}, \citenamefont {Machado}, \citenamefont {Vicuña-Hernández},
  \citenamefont {Picón}, \citenamefont {Celi}, \citenamefont {Torres},\ and\
  \citenamefont {Lewenstein}}]{pisanty_knotting_2019}%
  \BibitemOpen
  \bibfield  {author} {\bibinfo {author} {\bibfnamefont {E.}~\bibnamefont
  {Pisanty}}, \bibinfo {author} {\bibfnamefont {G.~J.}\ \bibnamefont
  {Machado}}, \bibinfo {author} {\bibfnamefont {V.}~\bibnamefont
  {Vicuña-Hernández}}, \bibinfo {author} {\bibfnamefont {A.}~\bibnamefont
  {Picón}}, \bibinfo {author} {\bibfnamefont {A.}~\bibnamefont {Celi}},
  \bibinfo {author} {\bibfnamefont {J.~P.}\ \bibnamefont {Torres}},\ and\
  \bibinfo {author} {\bibfnamefont {M.}~\bibnamefont {Lewenstein}},\ }\href
  {https://doi.org/10.1038/s41566-019-0450-2} {\bibfield  {journal} {\bibinfo
  {journal} {Nat. Photonics}\ }\textbf {\bibinfo {volume} {13}},\ \bibinfo
  {pages} {569} (\bibinfo {year} {2019})}\BibitemShut {NoStop}%
\bibitem [{\citenamefont {Kotlyar}\ \emph
  {et~al.}(2019{\natexlab{a}})\citenamefont {Kotlyar}, \citenamefont
  {Stafeev},\ and\ \citenamefont {Nalimov}}]{kotlyar_energy_2019}%
  \BibitemOpen
  \bibfield  {author} {\bibinfo {author} {\bibfnamefont {V.~V.}\ \bibnamefont
  {Kotlyar}}, \bibinfo {author} {\bibfnamefont {S.~S.}\ \bibnamefont
  {Stafeev}},\ and\ \bibinfo {author} {\bibfnamefont {A.~G.}\ \bibnamefont
  {Nalimov}},\ }\href {https://doi.org/10.1103/PhysRevA.99.033840} {\bibfield
  {journal} {\bibinfo  {journal} {Phys. Rev. A}\ }\textbf {\bibinfo {volume}
  {99}},\ \bibinfo {pages} {033840} (\bibinfo {year}
  {2019}{\natexlab{a}})}\BibitemShut {NoStop}%
\bibitem [{\citenamefont {Hancock}\ \emph {et~al.}(2019)\citenamefont
  {Hancock}, \citenamefont {Zahedpour}, \citenamefont {Goffin},\ and\
  \citenamefont {Milchberg}}]{hancock_free-space_2019}%
  \BibitemOpen
  \bibfield  {author} {\bibinfo {author} {\bibfnamefont {S.~W.}\ \bibnamefont
  {Hancock}}, \bibinfo {author} {\bibfnamefont {S.}~\bibnamefont {Zahedpour}},
  \bibinfo {author} {\bibfnamefont {A.}~\bibnamefont {Goffin}},\ and\ \bibinfo
  {author} {\bibfnamefont {H.~M.}\ \bibnamefont {Milchberg}},\ }\href
  {https://doi.org/10.1364/OPTICA.6.001547} {\bibfield  {journal} {\bibinfo
  {journal} {Optica}\ }\textbf {\bibinfo {volume} {6}},\ \bibinfo {pages}
  {1547} (\bibinfo {year} {2019})}\BibitemShut {NoStop}%
\bibitem [{\citenamefont {Bliokh}\ \emph {et~al.}(2019)\citenamefont {Bliokh},
  \citenamefont {Alonso},\ and\ \citenamefont
  {Dennis}}]{bliokh_geometric_2019}%
  \BibitemOpen
  \bibfield  {author} {\bibinfo {author} {\bibfnamefont {K.~Y.}\ \bibnamefont
  {Bliokh}}, \bibinfo {author} {\bibfnamefont {M.~A.}\ \bibnamefont {Alonso}},\
  and\ \bibinfo {author} {\bibfnamefont {M.~R.}\ \bibnamefont {Dennis}},\
  }\href {https://doi.org/10.1088/1361-6633/ab4415} {\bibfield  {journal}
  {\bibinfo  {journal} {Rep. Prog. Phys.}\ }\textbf {\bibinfo {volume} {82}},\
  \bibinfo {pages} {122401} (\bibinfo {year} {2019})}\BibitemShut {NoStop}%
\bibitem [{\citenamefont {Woźniak}\ \emph {et~al.}(2019)\citenamefont
  {Woźniak}, \citenamefont {Kurzynowski},\ and\ \citenamefont
  {Popiołek-Masajada}}]{wozniak_polarization_2019}%
  \BibitemOpen
  \bibfield  {author} {\bibinfo {author} {\bibfnamefont {W.~A.}\ \bibnamefont
  {Woźniak}}, \bibinfo {author} {\bibfnamefont {P.}~\bibnamefont
  {Kurzynowski}},\ and\ \bibinfo {author} {\bibfnamefont {A.}~\bibnamefont
  {Popiołek-Masajada}},\ }\href {https://doi.org/10.1016/j.optcom.2019.02.069}
  {\bibfield  {journal} {\bibinfo  {journal} {Opt. Commun.}\ }\textbf {\bibinfo
  {volume} {441}},\ \bibinfo {pages} {155} (\bibinfo {year}
  {2019})}\BibitemShut {NoStop}%
\bibitem [{\citenamefont {Chen}\ \emph
  {et~al.}(2019{\natexlab{a}})\citenamefont {Chen}, \citenamefont {Chen},\ and\
  \citenamefont {Liu}}]{chen_singularities_2019}%
  \BibitemOpen
  \bibfield  {author} {\bibinfo {author} {\bibfnamefont {W.}~\bibnamefont
  {Chen}}, \bibinfo {author} {\bibfnamefont {Y.}~\bibnamefont {Chen}},\ and\
  \bibinfo {author} {\bibfnamefont {W.}~\bibnamefont {Liu}},\ }\href
  {https://doi.org/10.1103/PhysRevLett.122.153907} {\bibfield  {journal}
  {\bibinfo  {journal} {Phys. Rev. Lett.}\ }\textbf {\bibinfo {volume} {122}},\
  \bibinfo {pages} {153907} (\bibinfo {year} {2019}{\natexlab{a}})}\BibitemShut
  {NoStop}%
\bibitem [{\citenamefont {Zhen}\ \emph {et~al.}(2014)\citenamefont {Zhen},
  \citenamefont {Hsu}, \citenamefont {Lu}, \citenamefont {Stone},\ and\
  \citenamefont {Soljačić}}]{zhen_topological_2014}%
  \BibitemOpen
  \bibfield  {author} {\bibinfo {author} {\bibfnamefont {B.}~\bibnamefont
  {Zhen}}, \bibinfo {author} {\bibfnamefont {C.~W.}\ \bibnamefont {Hsu}},
  \bibinfo {author} {\bibfnamefont {L.}~\bibnamefont {Lu}}, \bibinfo {author}
  {\bibfnamefont {A.~D.}\ \bibnamefont {Stone}},\ and\ \bibinfo {author}
  {\bibfnamefont {M.}~\bibnamefont {Soljačić}},\ }\href
  {https://doi.org/10.1103/PhysRevLett.113.257401} {\bibfield  {journal}
  {\bibinfo  {journal} {Phys. Rev. Lett.}\ }\textbf {\bibinfo {volume} {113}},\
  \bibinfo {pages} {257401} (\bibinfo {year} {2014})}\BibitemShut {NoStop}%
\bibitem [{\citenamefont {Zhang}\ \emph {et~al.}(2018)\citenamefont {Zhang},
  \citenamefont {Chen}, \citenamefont {Liu}, \citenamefont {Hsu}, \citenamefont
  {Wang}, \citenamefont {Guan}, \citenamefont {Liu}, \citenamefont {Shi},
  \citenamefont {Lu},\ and\ \citenamefont {Zi}}]{zhang_observation_2018}%
  \BibitemOpen
  \bibfield  {author} {\bibinfo {author} {\bibfnamefont {Y.}~\bibnamefont
  {Zhang}}, \bibinfo {author} {\bibfnamefont {A.}~\bibnamefont {Chen}},
  \bibinfo {author} {\bibfnamefont {W.}~\bibnamefont {Liu}}, \bibinfo {author}
  {\bibfnamefont {C.~W.}\ \bibnamefont {Hsu}}, \bibinfo {author} {\bibfnamefont
  {B.}~\bibnamefont {Wang}}, \bibinfo {author} {\bibfnamefont {F.}~\bibnamefont
  {Guan}}, \bibinfo {author} {\bibfnamefont {X.}~\bibnamefont {Liu}}, \bibinfo
  {author} {\bibfnamefont {L.}~\bibnamefont {Shi}}, \bibinfo {author}
  {\bibfnamefont {L.}~\bibnamefont {Lu}},\ and\ \bibinfo {author}
  {\bibfnamefont {J.}~\bibnamefont {Zi}},\ }\href
  {https://doi.org/10.1103/PhysRevLett.120.186103} {\bibfield  {journal}
  {\bibinfo  {journal} {Phys. Rev. Lett.}\ }\textbf {\bibinfo {volume} {120}},\
  \bibinfo {pages} {186103} (\bibinfo {year} {2018})}\BibitemShut {NoStop}%
\bibitem [{\citenamefont {Liu}\ \emph {et~al.}(2019)\citenamefont {Liu},
  \citenamefont {Wang}, \citenamefont {Zhang}, \citenamefont {Wang},
  \citenamefont {Zhao}, \citenamefont {Guan}, \citenamefont {Liu},
  \citenamefont {Shi},\ and\ \citenamefont {Zi}}]{liu_circularly_2019}%
  \BibitemOpen
  \bibfield  {author} {\bibinfo {author} {\bibfnamefont {W.}~\bibnamefont
  {Liu}}, \bibinfo {author} {\bibfnamefont {B.}~\bibnamefont {Wang}}, \bibinfo
  {author} {\bibfnamefont {Y.}~\bibnamefont {Zhang}}, \bibinfo {author}
  {\bibfnamefont {J.}~\bibnamefont {Wang}}, \bibinfo {author} {\bibfnamefont
  {M.}~\bibnamefont {Zhao}}, \bibinfo {author} {\bibfnamefont {F.}~\bibnamefont
  {Guan}}, \bibinfo {author} {\bibfnamefont {X.}~\bibnamefont {Liu}}, \bibinfo
  {author} {\bibfnamefont {L.}~\bibnamefont {Shi}},\ and\ \bibinfo {author}
  {\bibfnamefont {J.}~\bibnamefont {Zi}},\ }\href
  {https://doi.org/10.1103/PhysRevLett.123.116104} {\bibfield  {journal}
  {\bibinfo  {journal} {Phys. Rev. Lett.}\ }\textbf {\bibinfo {volume} {123}},\
  \bibinfo {pages} {116104} (\bibinfo {year} {2019})}\BibitemShut {NoStop}%
\bibitem [{\citenamefont {Chen}\ \emph
  {et~al.}(2019{\natexlab{b}})\citenamefont {Chen}, \citenamefont {Chen},\ and\
  \citenamefont {Liu}}]{PhysRevLett.122.153907}%
  \BibitemOpen
  \bibfield  {author} {\bibinfo {author} {\bibfnamefont {W.}~\bibnamefont
  {Chen}}, \bibinfo {author} {\bibfnamefont {Y.}~\bibnamefont {Chen}},\ and\
  \bibinfo {author} {\bibfnamefont {W.}~\bibnamefont {Liu}},\ }\href
  {https://doi.org/10.1103/PhysRevLett.122.153907} {\bibfield  {journal}
  {\bibinfo  {journal} {Phys. Rev. Lett.}\ }\textbf {\bibinfo {volume} {122}},\
  \bibinfo {pages} {153907} (\bibinfo {year} {2019}{\natexlab{b}})}\BibitemShut
  {NoStop}%
\bibitem [{\citenamefont {Chen}\ \emph
  {et~al.}(2020{\natexlab{a}})\citenamefont {Chen}, \citenamefont {Chen},\ and\
  \citenamefont {Liu}}]{doi:10.1002/lpor.202000049}%
  \BibitemOpen
  \bibfield  {author} {\bibinfo {author} {\bibfnamefont {W.}~\bibnamefont
  {Chen}}, \bibinfo {author} {\bibfnamefont {Y.}~\bibnamefont {Chen}},\ and\
  \bibinfo {author} {\bibfnamefont {W.}~\bibnamefont {Liu}},\ }\href
  {https://doi.org/10.1002/lpor.202000049} {\bibfield  {journal} {\bibinfo
  {journal} {Laser Photonics Rev.}\ }\textbf {\bibinfo {volume} {14}},\
  \bibinfo {pages} {2000049} (\bibinfo {year}
  {2020}{\natexlab{a}})}\BibitemShut {NoStop}%
\bibitem [{\citenamefont {Chen}\ \emph
  {et~al.}(2020{\natexlab{b}})\citenamefont {Chen}, \citenamefont {Yang},
  \citenamefont {Chen},\ and\ \citenamefont {Liu}}]{chen_global_2020}%
  \BibitemOpen
  \bibfield  {author} {\bibinfo {author} {\bibfnamefont {W.}~\bibnamefont
  {Chen}}, \bibinfo {author} {\bibfnamefont {Q.}~\bibnamefont {Yang}}, \bibinfo
  {author} {\bibfnamefont {Y.}~\bibnamefont {Chen}},\ and\ \bibinfo {author}
  {\bibfnamefont {W.}~\bibnamefont {Liu}},\ }\href
  {https://doi.org/10.1021/acsomega.0c01843} {\bibfield  {journal} {\bibinfo
  {journal} {ACS Omega}\ }\textbf {\bibinfo {volume} {5}},\ \bibinfo {pages}
  {14157} (\bibinfo {year} {2020}{\natexlab{b}})}\BibitemShut {NoStop}%
\bibitem [{\citenamefont {Zhou}\ \emph {et~al.}(2018)\citenamefont {Zhou},
  \citenamefont {Peng}, \citenamefont {Yoon}, \citenamefont {Hsu},
  \citenamefont {Nelson}, \citenamefont {Fu}, \citenamefont {Joannopoulos},
  \citenamefont {Solja{\v{c}}i{\'c}},\ and\ \citenamefont
  {Zhen}}]{noauthor_observation_nodate}%
  \BibitemOpen
  \bibfield  {author} {\bibinfo {author} {\bibfnamefont {H.}~\bibnamefont
  {Zhou}}, \bibinfo {author} {\bibfnamefont {C.}~\bibnamefont {Peng}}, \bibinfo
  {author} {\bibfnamefont {Y.}~\bibnamefont {Yoon}}, \bibinfo {author}
  {\bibfnamefont {C.~W.}\ \bibnamefont {Hsu}}, \bibinfo {author} {\bibfnamefont
  {K.~A.}\ \bibnamefont {Nelson}}, \bibinfo {author} {\bibfnamefont
  {L.}~\bibnamefont {Fu}}, \bibinfo {author} {\bibfnamefont {J.~D.}\
  \bibnamefont {Joannopoulos}}, \bibinfo {author} {\bibfnamefont
  {M.}~\bibnamefont {Solja{\v{c}}i{\'c}}},\ and\ \bibinfo {author}
  {\bibfnamefont {B.}~\bibnamefont {Zhen}},\ }\href
  {https://doi.org/10.1126/science.aap9859} {\bibfield  {journal} {\bibinfo
  {journal} {Science}\ }\textbf {\bibinfo {volume} {359}},\ \bibinfo {pages}
  {1009} (\bibinfo {year} {2018})}\BibitemShut {NoStop}%
\bibitem [{\citenamefont {Fösel}\ \emph {et~al.}(2017)\citenamefont {Fösel},
  \citenamefont {Peano},\ and\ \citenamefont {Marquardt}}]{fosel_l_2017}%
  \BibitemOpen
  \bibfield  {author} {\bibinfo {author} {\bibfnamefont {T.}~\bibnamefont
  {Fösel}}, \bibinfo {author} {\bibfnamefont {V.}~\bibnamefont {Peano}},\ and\
  \bibinfo {author} {\bibfnamefont {F.}~\bibnamefont {Marquardt}},\ }\href
  {https://doi.org/10.1088/1367-2630/aa8a9f} {\bibfield  {journal} {\bibinfo
  {journal} {New J. Phys.}\ }\textbf {\bibinfo {volume} {19}},\ \bibinfo
  {pages} {115013} (\bibinfo {year} {2017})}\BibitemShut {NoStop}%
\bibitem [{\citenamefont {Dennis}(2002)}]{dennis_polarization_2002}%
  \BibitemOpen
  \bibfield  {author} {\bibinfo {author} {\bibfnamefont {M.}~\bibnamefont
  {Dennis}},\ }\href {https://doi.org/10.1016/S0030-4018(02)02088-6} {\bibfield
   {journal} {\bibinfo  {journal} {Opt. Commun.}\ }\textbf {\bibinfo {volume}
  {213}},\ \bibinfo {pages} {201} (\bibinfo {year} {2002})}\BibitemShut
  {NoStop}%
\bibitem [{\citenamefont {Berry}(2004{\natexlab{a}})}]{berry_electric_2004}%
  \BibitemOpen
  \bibfield  {author} {\bibinfo {author} {\bibfnamefont {M.~V.}\ \bibnamefont
  {Berry}},\ }\href {https://doi.org/10.1088/1464-4258/6/5/030} {\bibfield
  {journal} {\bibinfo  {journal} {J. Opt. A: Pure Appl. Opt.}\ }\textbf
  {\bibinfo {volume} {6}},\ \bibinfo {pages} {475} (\bibinfo {year}
  {2004}{\natexlab{a}})}\BibitemShut {NoStop}%
\bibitem [{\citenamefont {Freund}(2002)}]{freund_polarization_2002}%
  \BibitemOpen
  \bibfield  {author} {\bibinfo {author} {\bibfnamefont {I.}~\bibnamefont
  {Freund}},\ }\href {https://doi.org/10.1016/S0030-4018(01)01725-4} {\bibfield
   {journal} {\bibinfo  {journal} {Opt. Commun.}\ }\textbf {\bibinfo {volume}
  {201}},\ \bibinfo {pages} {251} (\bibinfo {year} {2002})}\BibitemShut
  {NoStop}%
\bibitem [{\citenamefont {Berry}(2004{\natexlab{b}})}]{berry_index_2004}%
  \BibitemOpen
  \bibfield  {author} {\bibinfo {author} {\bibfnamefont {M.~V.}\ \bibnamefont
  {Berry}},\ }\href {https://doi.org/10.1088/1464-4258/6/7/003} {\bibfield
  {journal} {\bibinfo  {journal} {J. Opt. A: Pure Appl. Opt.}\ }\textbf
  {\bibinfo {volume} {6}},\ \bibinfo {pages} {675} (\bibinfo {year}
  {2004}{\natexlab{b}})}\BibitemShut {NoStop}%
\bibitem [{\citenamefont {Angelsky}\ \emph {et~al.}(2002)\citenamefont
  {Angelsky}, \citenamefont {Mokhun}, \citenamefont {Mokhun},\ and\
  \citenamefont {Soskin}}]{angelsky_interferometric_2002}%
  \BibitemOpen
  \bibfield  {author} {\bibinfo {author} {\bibfnamefont {O.~V.}\ \bibnamefont
  {Angelsky}}, \bibinfo {author} {\bibfnamefont {I.~I.}\ \bibnamefont
  {Mokhun}}, \bibinfo {author} {\bibfnamefont {A.~I.}\ \bibnamefont {Mokhun}},\
  and\ \bibinfo {author} {\bibfnamefont {M.~S.}\ \bibnamefont {Soskin}},\
  }\href {https://doi.org/10.1103/PhysRevE.65.036602} {\bibfield  {journal}
  {\bibinfo  {journal} {Phys. Rev. E}\ }\textbf {\bibinfo {volume} {65}},\
  \bibinfo {pages} {036602} (\bibinfo {year} {2002})}\BibitemShut {NoStop}%
\bibitem [{\citenamefont {Flossmann}\ \emph {et~al.}(2005)\citenamefont
  {Flossmann}, \citenamefont {Schwarz}, \citenamefont {Maier},\ and\
  \citenamefont {Dennis}}]{flossmann_polarization_2005}%
  \BibitemOpen
  \bibfield  {author} {\bibinfo {author} {\bibfnamefont {F.}~\bibnamefont
  {Flossmann}}, \bibinfo {author} {\bibfnamefont {U.~T.}\ \bibnamefont
  {Schwarz}}, \bibinfo {author} {\bibfnamefont {M.}~\bibnamefont {Maier}},\
  and\ \bibinfo {author} {\bibfnamefont {M.~R.}\ \bibnamefont {Dennis}},\
  }\href {https://doi.org/10.1103/PhysRevLett.95.253901} {\bibfield  {journal}
  {\bibinfo  {journal} {Phys. Rev. Lett.}\ }\textbf {\bibinfo {volume} {95}},\
  \bibinfo {pages} {253901} (\bibinfo {year} {2005})}\BibitemShut {NoStop}%
\bibitem [{\citenamefont {Flossmann}\ \emph {et~al.}(2008)\citenamefont
  {Flossmann}, \citenamefont {O‘Holleran}, \citenamefont {Dennis},\ and\
  \citenamefont {Padgett}}]{flossmann_polarization_2008}%
  \BibitemOpen
  \bibfield  {author} {\bibinfo {author} {\bibfnamefont {F.}~\bibnamefont
  {Flossmann}}, \bibinfo {author} {\bibfnamefont {K.}~\bibnamefont
  {O‘Holleran}}, \bibinfo {author} {\bibfnamefont {M.~R.}\ \bibnamefont
  {Dennis}},\ and\ \bibinfo {author} {\bibfnamefont {M.~J.}\ \bibnamefont
  {Padgett}},\ }\href {https://doi.org/10.1103/PhysRevLett.100.203902}
  {\bibfield  {journal} {\bibinfo  {journal} {Phys. Rev. Lett.}\ }\textbf
  {\bibinfo {volume} {100}},\ \bibinfo {pages} {203902} (\bibinfo {year}
  {2008})}\BibitemShut {NoStop}%
\bibitem [{\citenamefont {Burresi}\ \emph {et~al.}(2009)\citenamefont
  {Burresi}, \citenamefont {Engelen}, \citenamefont {Opheij}, \citenamefont
  {van Oosten}, \citenamefont {Mori}, \citenamefont {Baba},\ and\ \citenamefont
  {Kuipers}}]{burresi_observation_2009}%
  \BibitemOpen
  \bibfield  {author} {\bibinfo {author} {\bibfnamefont {M.}~\bibnamefont
  {Burresi}}, \bibinfo {author} {\bibfnamefont {R.~J.~P.}\ \bibnamefont
  {Engelen}}, \bibinfo {author} {\bibfnamefont {A.}~\bibnamefont {Opheij}},
  \bibinfo {author} {\bibfnamefont {D.}~\bibnamefont {van Oosten}}, \bibinfo
  {author} {\bibfnamefont {D.}~\bibnamefont {Mori}}, \bibinfo {author}
  {\bibfnamefont {T.}~\bibnamefont {Baba}},\ and\ \bibinfo {author}
  {\bibfnamefont {L.}~\bibnamefont {Kuipers}},\ }\href
  {https://doi.org/10.1103/PhysRevLett.102.033902} {\bibfield  {journal}
  {\bibinfo  {journal} {Phys. Rev. Lett.}\ }\textbf {\bibinfo {volume} {102}},\
  \bibinfo {pages} {033902} (\bibinfo {year} {2009})}\BibitemShut {NoStop}%
\bibitem [{\citenamefont {Kurzynowski}\ \emph {et~al.}(2010)\citenamefont
  {Kurzynowski}, \citenamefont {Woźniak},\ and\ \citenamefont
  {Borwińska}}]{kurzynowski_regular_2010}%
  \BibitemOpen
  \bibfield  {author} {\bibinfo {author} {\bibfnamefont {P.}~\bibnamefont
  {Kurzynowski}}, \bibinfo {author} {\bibfnamefont {W.~A.}\ \bibnamefont
  {Woźniak}},\ and\ \bibinfo {author} {\bibfnamefont {M.}~\bibnamefont
  {Borwińska}},\ }\href {https://doi.org/10.1088/2040-8978/12/3/035406}
  {\bibfield  {journal} {\bibinfo  {journal} {J. Opt.}\ }\textbf {\bibinfo
  {volume} {12}},\ \bibinfo {pages} {035406} (\bibinfo {year}
  {2010})}\BibitemShut {NoStop}%
\bibitem [{\citenamefont {Park}\ \emph {et~al.}(2010)\citenamefont {Park},
  \citenamefont {Lee},\ and\ \citenamefont {Lee}}]{park_polarization_2010}%
  \BibitemOpen
  \bibfield  {author} {\bibinfo {author} {\bibfnamefont {J.}~\bibnamefont
  {Park}}, \bibinfo {author} {\bibfnamefont {S.-Y.}\ \bibnamefont {Lee}},\ and\
  \bibinfo {author} {\bibfnamefont {B.}~\bibnamefont {Lee}},\ }\href
  {https://doi.org/10.1109/JQE.2010.2053699} {\bibfield  {journal} {\bibinfo
  {journal} {IEEE J. Quantum Electron.}\ }\textbf {\bibinfo {volume} {46}},\
  \bibinfo {pages} {1577} (\bibinfo {year} {2010})}\BibitemShut {NoStop}%
\bibitem [{\citenamefont {Vyas}\ \emph {et~al.}(2013)\citenamefont {Vyas},
  \citenamefont {Kozawa},\ and\ \citenamefont {Sato}}]{vyas_polarization_2013}%
  \BibitemOpen
  \bibfield  {author} {\bibinfo {author} {\bibfnamefont {S.}~\bibnamefont
  {Vyas}}, \bibinfo {author} {\bibfnamefont {Y.}~\bibnamefont {Kozawa}},\ and\
  \bibinfo {author} {\bibfnamefont {S.}~\bibnamefont {Sato}},\ }\href
  {https://doi.org/10.1364/OE.21.008972} {\bibfield  {journal} {\bibinfo
  {journal} {Opt. Express}\ }\textbf {\bibinfo {volume} {21}},\ \bibinfo
  {pages} {8972} (\bibinfo {year} {2013})}\BibitemShut {NoStop}%
\bibitem [{\citenamefont {Bauer}\ \emph {et~al.}(2015)\citenamefont {Bauer},
  \citenamefont {Banzer}, \citenamefont {Karimi}, \citenamefont {Orlov},
  \citenamefont {Rubano}, \citenamefont {Marrucci}, \citenamefont {Santamato},
  \citenamefont {Boyd},\ and\ \citenamefont {Leuchs}}]{bauer_observation_2015}%
  \BibitemOpen
  \bibfield  {author} {\bibinfo {author} {\bibfnamefont {T.}~\bibnamefont
  {Bauer}}, \bibinfo {author} {\bibfnamefont {P.}~\bibnamefont {Banzer}},
  \bibinfo {author} {\bibfnamefont {E.}~\bibnamefont {Karimi}}, \bibinfo
  {author} {\bibfnamefont {S.}~\bibnamefont {Orlov}}, \bibinfo {author}
  {\bibfnamefont {A.}~\bibnamefont {Rubano}}, \bibinfo {author} {\bibfnamefont
  {L.}~\bibnamefont {Marrucci}}, \bibinfo {author} {\bibfnamefont
  {E.}~\bibnamefont {Santamato}}, \bibinfo {author} {\bibfnamefont {R.~W.}\
  \bibnamefont {Boyd}},\ and\ \bibinfo {author} {\bibfnamefont
  {G.}~\bibnamefont {Leuchs}},\ }\href
  {https://doi.org/10.1126/science.1260635} {\bibfield  {journal} {\bibinfo
  {journal} {Science}\ }\textbf {\bibinfo {volume} {347}},\ \bibinfo {pages}
  {964} (\bibinfo {year} {2015})}\BibitemShut {NoStop}%
\bibitem [{\citenamefont {Ye}\ \emph {et~al.}(2017)\citenamefont {Ye},
  \citenamefont {Peng}, \citenamefont {Zhou}, \citenamefont {Xin},\ and\
  \citenamefont {Song}}]{ye_simulation_2017}%
  \BibitemOpen
  \bibfield  {author} {\bibinfo {author} {\bibfnamefont {D.}~\bibnamefont
  {Ye}}, \bibinfo {author} {\bibfnamefont {X.}~\bibnamefont {Peng}}, \bibinfo
  {author} {\bibfnamefont {M.}~\bibnamefont {Zhou}}, \bibinfo {author}
  {\bibfnamefont {Y.}~\bibnamefont {Xin}},\ and\ \bibinfo {author}
  {\bibfnamefont {M.}~\bibnamefont {Song}},\ }\href
  {https://doi.org/10.1364/JOSAA.34.001957} {\bibfield  {journal} {\bibinfo
  {journal} {J. Opt. Soc. Am. A}\ }\textbf {\bibinfo {volume} {34}},\ \bibinfo
  {pages} {1957} (\bibinfo {year} {2017})}\BibitemShut {NoStop}%
\bibitem [{\citenamefont {Kotlyar}\ \emph
  {et~al.}(2019{\natexlab{b}})\citenamefont {Kotlyar}, \citenamefont {Nalimov},
  \citenamefont {Stafeev},\ and\ \citenamefont
  {O’Faolain}}]{kotlyar_single_2019}%
  \BibitemOpen
  \bibfield  {author} {\bibinfo {author} {\bibfnamefont {V.~V.}\ \bibnamefont
  {Kotlyar}}, \bibinfo {author} {\bibfnamefont {A.~G.}\ \bibnamefont
  {Nalimov}}, \bibinfo {author} {\bibfnamefont {S.~S.}\ \bibnamefont
  {Stafeev}},\ and\ \bibinfo {author} {\bibfnamefont {L.}~\bibnamefont
  {O’Faolain}},\ }\href {https://doi.org/10.1088/2040-8986/ab14c8} {\bibfield
   {journal} {\bibinfo  {journal} {J. Opt.}\ }\textbf {\bibinfo {volume}
  {21}},\ \bibinfo {pages} {055004} (\bibinfo {year}
  {2019}{\natexlab{b}})}\BibitemShut {NoStop}%
\bibitem [{\citenamefont {Chen}\ \emph
  {et~al.}(2019{\natexlab{c}})\citenamefont {Chen}, \citenamefont {Liu},
  \citenamefont {Zhang}, \citenamefont {Wang}, \citenamefont {Liu},
  \citenamefont {Shi}, \citenamefont {Lu},\ and\ \citenamefont
  {Zi}}]{chen_observing_2019}%
  \BibitemOpen
  \bibfield  {author} {\bibinfo {author} {\bibfnamefont {A.}~\bibnamefont
  {Chen}}, \bibinfo {author} {\bibfnamefont {W.}~\bibnamefont {Liu}}, \bibinfo
  {author} {\bibfnamefont {Y.}~\bibnamefont {Zhang}}, \bibinfo {author}
  {\bibfnamefont {B.}~\bibnamefont {Wang}}, \bibinfo {author} {\bibfnamefont
  {X.}~\bibnamefont {Liu}}, \bibinfo {author} {\bibfnamefont {L.}~\bibnamefont
  {Shi}}, \bibinfo {author} {\bibfnamefont {L.}~\bibnamefont {Lu}},\ and\
  \bibinfo {author} {\bibfnamefont {J.}~\bibnamefont {Zi}},\ }\href
  {https://doi.org/10.1103/PhysRevB.99.180101} {\bibfield  {journal} {\bibinfo
  {journal} {Phys. Rev. B}\ }\textbf {\bibinfo {volume} {99}},\ \bibinfo
  {pages} {180101} (\bibinfo {year} {2019}{\natexlab{c}})}\BibitemShut
  {NoStop}%
\bibitem [{\citenamefont {Pal}\ \emph {et~al.}(2017{\natexlab{a}})\citenamefont
  {Pal}, \citenamefont {{Ruchi}},\ and\ \citenamefont
  {Senthilkumaran}}]{kumar_pal_c-point_2017}%
  \BibitemOpen
  \bibfield  {author} {\bibinfo {author} {\bibfnamefont {S.~K.}\ \bibnamefont
  {Pal}}, \bibinfo {author} {\bibnamefont {{Ruchi}}},\ and\ \bibinfo {author}
  {\bibfnamefont {P.}~\bibnamefont {Senthilkumaran}},\ }\href
  {https://doi.org/10.1016/j.optcom.2017.02.048} {\bibfield  {journal}
  {\bibinfo  {journal} {Opt. Commun.}\ }\textbf {\bibinfo {volume} {393}},\
  \bibinfo {pages} {156} (\bibinfo {year} {2017}{\natexlab{a}})}\BibitemShut
  {NoStop}%
\bibitem [{\citenamefont {Pal}\ \emph {et~al.}(2017{\natexlab{b}})\citenamefont
  {Pal}, \citenamefont {{Ruchi}},\ and\ \citenamefont
  {Senthilkumaran}}]{pal_polarization_2017}%
  \BibitemOpen
  \bibfield  {author} {\bibinfo {author} {\bibfnamefont {S.~K.}\ \bibnamefont
  {Pal}}, \bibinfo {author} {\bibnamefont {{Ruchi}}},\ and\ \bibinfo {author}
  {\bibfnamefont {P.}~\bibnamefont {Senthilkumaran}},\ }\href
  {https://doi.org/10.1364/AO.56.006181} {\bibfield  {journal} {\bibinfo
  {journal} {Appl. Opt.}\ }\textbf {\bibinfo {volume} {56}},\ \bibinfo {pages}
  {6181} (\bibinfo {year} {2017}{\natexlab{b}})}\BibitemShut {NoStop}%
\bibitem [{\citenamefont {Garcia-Etxarri}(2017)}]{garcia-etxarri_optical_2017}%
  \BibitemOpen
  \bibfield  {author} {\bibinfo {author} {\bibfnamefont {A.}~\bibnamefont
  {Garcia-Etxarri}},\ }\href {https://doi.org/10.1021/acsphotonics.7b00002}
  {\bibfield  {journal} {\bibinfo  {journal} {ACS Photonics}\ }\textbf
  {\bibinfo {volume} {4}},\ \bibinfo {pages} {1159} (\bibinfo {year}
  {2017})}\BibitemShut {NoStop}%
\bibitem [{\citenamefont {Grigoriev}\ \emph {et~al.}(2018)\citenamefont
  {Grigoriev}, \citenamefont {Kuznetsov}, \citenamefont {Vladimirova},\ and\
  \citenamefont {Makarov}}]{PhysRevA.98.063805}%
  \BibitemOpen
  \bibfield  {author} {\bibinfo {author} {\bibfnamefont {K.~S.}\ \bibnamefont
  {Grigoriev}}, \bibinfo {author} {\bibfnamefont {N.~Y.}\ \bibnamefont
  {Kuznetsov}}, \bibinfo {author} {\bibfnamefont {Y.~V.}\ \bibnamefont
  {Vladimirova}},\ and\ \bibinfo {author} {\bibfnamefont {V.~A.}\ \bibnamefont
  {Makarov}},\ }\href {https://doi.org/10.1103/PhysRevA.98.063805} {\bibfield
  {journal} {\bibinfo  {journal} {Phys. Rev. A}\ }\textbf {\bibinfo {volume}
  {98}},\ \bibinfo {pages} {063805} (\bibinfo {year} {2018})}\BibitemShut
  {NoStop}%
\bibitem [{\citenamefont {Kuznetsov}\ \emph {et~al.}(2020)\citenamefont
  {Kuznetsov}, \citenamefont {Grigoriev}, \citenamefont {Vladimirova},\ and\
  \citenamefont {Makarov}}]{YuKuznetsov:20}%
  \BibitemOpen
  \bibfield  {author} {\bibinfo {author} {\bibfnamefont {N.~Y.}\ \bibnamefont
  {Kuznetsov}}, \bibinfo {author} {\bibfnamefont {K.~S.}\ \bibnamefont
  {Grigoriev}}, \bibinfo {author} {\bibfnamefont {Y.~V.}\ \bibnamefont
  {Vladimirova}},\ and\ \bibinfo {author} {\bibfnamefont {V.~A.}\ \bibnamefont
  {Makarov}},\ }\href {https://doi.org/10.1364/OE.398602} {\bibfield  {journal}
  {\bibinfo  {journal} {Opt. Express}\ }\textbf {\bibinfo {volume} {28}},\
  \bibinfo {pages} {27293} (\bibinfo {year} {2020})}\BibitemShut {NoStop}%
\bibitem [{\citenamefont {Berry}(2001)}]{berry_geometry_n.d.}%
  \BibitemOpen
  \bibfield  {author} {\bibinfo {author} {\bibfnamefont {M.~V.}\ \bibnamefont
  {Berry}},\ }\href {https://doi.org/10.1117/12.428252} {\emph {\bibinfo
  {title} {Second International Conference on Singular Optics (Optical
  Vortices): Fundamentals and Applications}}},\ Vol.\ \bibinfo {volume} {4403}\
  (\bibinfo  {publisher} {SPIE},\ \bibinfo {year} {2001})\BibitemShut {NoStop}%
\bibitem [{\citenamefont {Berry}(1984)}]{quantal_phase_n.d.}%
  \BibitemOpen
  \bibfield  {author} {\bibinfo {author} {\bibfnamefont {M.~V.}\ \bibnamefont
  {Berry}},\ }\href {https://doi.org/https://doi.org/10.1098/rspa.1984.0023}
  {\bibfield  {journal} {\bibinfo  {journal} {Proc. R. Soc. London. A}\
  }\textbf {\bibinfo {volume} {392}},\ \bibinfo {pages} {45} (\bibinfo {year}
  {1984})}\BibitemShut {NoStop}%
\bibitem [{\citenamefont {Shen}\ \emph {et~al.}(2019)\citenamefont {Shen},
  \citenamefont {Wang}, \citenamefont {Xie}, \citenamefont {Min}, \citenamefont
  {Fu}, \citenamefont {Liu}, \citenamefont {Gong},\ and\ \citenamefont
  {Yuan}}]{shen_optical_2019}%
  \BibitemOpen
  \bibfield  {author} {\bibinfo {author} {\bibfnamefont {Y.}~\bibnamefont
  {Shen}}, \bibinfo {author} {\bibfnamefont {X.}~\bibnamefont {Wang}}, \bibinfo
  {author} {\bibfnamefont {Z.}~\bibnamefont {Xie}}, \bibinfo {author}
  {\bibfnamefont {C.}~\bibnamefont {Min}}, \bibinfo {author} {\bibfnamefont
  {X.}~\bibnamefont {Fu}}, \bibinfo {author} {\bibfnamefont {Q.}~\bibnamefont
  {Liu}}, \bibinfo {author} {\bibfnamefont {M.}~\bibnamefont {Gong}},\ and\
  \bibinfo {author} {\bibfnamefont {X.}~\bibnamefont {Yuan}},\ }\href
  {https://doi.org/10.1038/s41377-019-0194-2} {\bibfield  {journal} {\bibinfo
  {journal} {Light: Sci. Appl.}\ }\textbf {\bibinfo {volume} {8}},\ \bibinfo
  {pages} {90} (\bibinfo {year} {2019})}\BibitemShut {NoStop}%
\bibitem [{\citenamefont {Ye}\ \emph {et~al.}(2020)\citenamefont {Ye},
  \citenamefont {Gao},\ and\ \citenamefont {Liu}}]{ye_singular_2020}%
  \BibitemOpen
  \bibfield  {author} {\bibinfo {author} {\bibfnamefont {W.}~\bibnamefont
  {Ye}}, \bibinfo {author} {\bibfnamefont {Y.}~\bibnamefont {Gao}},\ and\
  \bibinfo {author} {\bibfnamefont {J.}~\bibnamefont {Liu}},\ }\href
  {https://doi.org/10.1103/PhysRevLett.124.153904} {\bibfield  {journal}
  {\bibinfo  {journal} {Phys. Rev. Lett.}\ }\textbf {\bibinfo {volume} {124}},\
  \bibinfo {pages} {153904} (\bibinfo {year} {2020})}\BibitemShut {NoStop}%
\bibitem [{COM()}]{COMSOL}%
  \BibitemOpen
  \href@noop {} {}\bibinfo {howpublished} {\url{www.comsol.com}}\BibitemShut
  {NoStop}%
\bibitem [{\citenamefont {Rose}(1955)}]{Rose_Multipole_1955}%
  \BibitemOpen
  \bibfield  {author} {\bibinfo {author} {\bibfnamefont {M.~E.}\ \bibnamefont
  {Rose}},\ }\href@noop {} {\emph {\bibinfo {title} {Multipole Fields}}}\
  (\bibinfo  {publisher} {John Wiley},\ \bibinfo {address} {New York},\
  \bibinfo {year} {1955})\BibitemShut {NoStop}%
\bibitem [{\citenamefont {Xu}(1995)}]{xu_electromagnetic_n.d.}%
  \BibitemOpen
  \bibfield  {author} {\bibinfo {author} {\bibfnamefont {Y.}~\bibnamefont
  {Xu}},\ }\href {https://doi.org/10.1364/AO.34.004573} {\bibfield  {journal}
  {\bibinfo  {journal} {Appl. Opt.}\ }\textbf {\bibinfo {volume} {34}},\
  \bibinfo {pages} {4573} (\bibinfo {year} {1995})}\BibitemShut {NoStop}%
\bibitem [{\citenamefont {Bohren}\ and\ \citenamefont
  {Huffman}(2004)}]{bohren_absorption_2004}%
  \BibitemOpen
  \bibfield  {author} {\bibinfo {author} {\bibfnamefont {C.~F.}\ \bibnamefont
  {Bohren}}\ and\ \bibinfo {author} {\bibfnamefont {D.~R.}\ \bibnamefont
  {Huffman}},\ }\href@noop {} {\emph {\bibinfo {title} {Absorption and
  Scattering of Light by Small Particles}}}\ (\bibinfo  {publisher}
  {Wiley-VCH},\ \bibinfo {address} {Weinheim},\ \bibinfo {year}
  {2004})\BibitemShut {NoStop}%
\bibitem [{\citenamefont {Ruppin}(2006)}]{ruppin_scattering_2006}%
  \BibitemOpen
  \bibfield  {author} {\bibinfo {author} {\bibfnamefont {R.}~\bibnamefont
  {Ruppin}},\ }\href {https://doi.org/10.1163/156939306779292390} {\bibfield
  {journal} {\bibinfo  {journal} {J. Electromagn. Wav. Appl.}\ }\textbf
  {\bibinfo {volume} {20}},\ \bibinfo {pages} {1569} (\bibinfo {year}
  {2006})}\BibitemShut {NoStop}%
\bibitem [{\citenamefont {Frankel}(2012)}]{frankel_geometry_2012}%
  \BibitemOpen
  \bibfield  {author} {\bibinfo {author} {\bibfnamefont {T.}~\bibnamefont
  {Frankel}},\ }\href@noop {} {\emph {\bibinfo {title} {The Geometry of
  Physics: An Introduction}}},\ \bibinfo {edition} {3rd}\ ed.\ (\bibinfo
  {publisher} {Cambridge University Press},\ \bibinfo {address} {Cambridge ;
  New York},\ \bibinfo {year} {2012})\BibitemShut {NoStop}%
\bibitem [{\citenamefont {Berry}(2013)}]{Berry_2013}%
  \BibitemOpen
  \bibfield  {author} {\bibinfo {author} {\bibfnamefont {M.~V.}\ \bibnamefont
  {Berry}},\ }\href {https://doi.org/10.1088/2040-8978/15/4/044024} {\bibfield
  {journal} {\bibinfo  {journal} {J. Opt.}\ }\textbf {\bibinfo {volume} {15}},\
  \bibinfo {pages} {044024} (\bibinfo {year} {2013})}\BibitemShut {NoStop}%
\bibitem [{\citenamefont {Wang}\ and\ \citenamefont
  {Chan}(2014)}]{wang_lateral_2014}%
  \BibitemOpen
  \bibfield  {author} {\bibinfo {author} {\bibfnamefont {S.~B.}\ \bibnamefont
  {Wang}}\ and\ \bibinfo {author} {\bibfnamefont {C.~T.}\ \bibnamefont
  {Chan}},\ }\href {https://doi.org/10.1038/ncomms4307} {\bibfield  {journal}
  {\bibinfo  {journal} {Nat. Commun.}\ }\textbf {\bibinfo {volume} {5}},\
  \bibinfo {pages} {3307} (\bibinfo {year} {2014})}\BibitemShut {NoStop}%
\bibitem [{\citenamefont {Bliokh}\ \emph {et~al.}(2015)\citenamefont {Bliokh},
  \citenamefont {Rodríguez-Fortuño}, \citenamefont {Nori},\ and\
  \citenamefont {Zayats}}]{bliokh_spinorbit_2015}%
  \BibitemOpen
  \bibfield  {author} {\bibinfo {author} {\bibfnamefont {K.~Y.}\ \bibnamefont
  {Bliokh}}, \bibinfo {author} {\bibfnamefont {F.~J.}\ \bibnamefont
  {Rodríguez-Fortuño}}, \bibinfo {author} {\bibfnamefont {F.}~\bibnamefont
  {Nori}},\ and\ \bibinfo {author} {\bibfnamefont {A.~V.}\ \bibnamefont
  {Zayats}},\ }\href {https://doi.org/10.1038/nphoton.2015.201} {\bibfield
  {journal} {\bibinfo  {journal} {Nat. Photonics}\ }\textbf {\bibinfo {volume}
  {9}},\ \bibinfo {pages} {796} (\bibinfo {year} {2015})}\BibitemShut {NoStop}%
\bibitem [{\citenamefont {Wang}\ \emph {et~al.}(2019)\citenamefont {Wang},
  \citenamefont {Hou}, \citenamefont {Lu}, \citenamefont {Chen}, \citenamefont
  {Zhang},\ and\ \citenamefont {Chan}}]{wang2019arbitrary}%
  \BibitemOpen
  \bibfield  {author} {\bibinfo {author} {\bibfnamefont {S.}~\bibnamefont
  {Wang}}, \bibinfo {author} {\bibfnamefont {B.}~\bibnamefont {Hou}}, \bibinfo
  {author} {\bibfnamefont {W.}~\bibnamefont {Lu}}, \bibinfo {author}
  {\bibfnamefont {Y.}~\bibnamefont {Chen}}, \bibinfo {author} {\bibfnamefont
  {Z.}~\bibnamefont {Zhang}},\ and\ \bibinfo {author} {\bibfnamefont {C.~T.}\
  \bibnamefont {Chan}},\ }\href {https://doi.org/10.1038/s41467-019-08826-6}
  {\bibfield  {journal} {\bibinfo  {journal} {Nat. Commun.}\ }\textbf {\bibinfo
  {volume} {10}},\ \bibinfo {pages} {832} (\bibinfo {year} {2019})}\BibitemShut
  {NoStop}%
\end{thebibliography}%
\end{document}